# Uncertainties in the attribution of greenhouse gas warming and implications for climate prediction

Gareth S. Jones*†, Peter A. Stott†and John F. B. Mitchell†

†- Met Office Hadley Centre, Exeter, UK

## Abstract

Using optimal detection techniques with climate model simulations, most of the observed increase of near surface temperatures over the second half of the twentieth century is attributed to anthropogenic influences. However, the partitioning of the anthropogenic influence to individual factors, such as greenhouse gases and aerosols, is much less robust. Differences in how forcing factors are applied, in their radiative influence and in models' climate sensitivities, substantially influence the response patterns. We find standard optimal detection methodologies cannot fully reconcile this response diversity. By selecting a set of experiments to enable the diagnosing of greenhouse gases and the combined influence of other anthropogenic and natural factors, we find robust detections of well mixed greenhouse gases across a large ensemble of models. Of the observed warming over the 20th century of 0.65K/century we find, using a multi model mean not incorporating pattern uncertainty, a well mixed greenhouse gas warming of 0.87 to 1.22K/century. This is partially offset by cooling from other anthropogenic and natural influences of -0.54 to -0.22K/century. Although better constrained than recent studies, the attributable trends across climate models are still wide, with implications for observational constrained estimates of transient climate response. Some of the uncertainties could be reduced in future by having more model data to better quantify the simulated estimates of the signals and natural variability, by designing model experiments more effectively and better quantification of the climate model radiative influences. Most importantly, how model pattern uncertainties are incorporated into the optimal detection methodology should be improved.

## 1    Introduction

Detection of climate change and attribution of its causes are important for the understanding of climate change, for evaluating climate models and for helping to constrain predictions of anthropogenic climate change. Formal detection studies have examined observed climate changes over a variety of time and spatial scales (see for example, *Bindoff et al.*, 2013). These analyses compare observations with spatio-temporal patterns produced from climate models that represent the responses to different climate forcing factors, such as well mixed greenhouse gases, aerosols, solar irradiance and volcanic aerosol influences. Measures of internal variability are incorporated into sophisticated statistical analysis to attempt to detect the influence of the different forcing factors over internal climate variability.

The use of spatio-temporal patterns of temperature in detection studies has been reported to improve the consistency of the attribution of greenhouse gas warming across different models (*Stott et al.*, 2006). Such an approach attempts to compensate for gross errors in the models climate sensitivity and in the applied radiative forcings by scaling up under responsive models and scaling down over responsive models to better match the observations (*Allen et al.*, 2000). Using the technique has also been used to constrain 21st century temperature predictions (*Allen et al.*, 2000; *Stott and Kettleborough*, 2002; *Kettleborough et al.*, 2007), with the result of bringing closer together different model projections of future temperature change than if observations were not used as a constraint. This type of agreement across models, with results predominantly sensitive only to the observations, has been termed 'Stable Inference from Data' (*Stott et al.*, 2006; *Kettleborough et al.*, 2007).

However, this viewpoint has been somewhat challenged by results from recent detection studies. *Gillett et al.* (2013); *Jones et al.* (2013) and *Ribes and Terray* (2013) have taken advantage of large numbers of climate model simulations made available from the Coupled Model Intercomparison Project (CMIP5) (*Taylor et al.*, 2012) to deduce the contribution to changes in near surface temperatures over the last 60-150 years. These studies concluded that attributed trends for greenhouse gases over the last 60 or so years vary substantially, depending on which climate models were used in the detection analyses (Figure 10.4 in *Bindoff et al.*, 2013). This sensitivity of results to model used could indicate that observationally constrained estimates of 21st century temperatures may not be as robust as previously thought.

---

*Corresponding author:Gareth S Jones, Met Office Hadley Centre, FitzRoy Road, Exeter, EX1 3PB, UK (gareth.s.jones@metoffice.gov.uk)

There have been a couple of ways of potentially dealing with this criticism. First, it has been argued that using a multi-model average gives a more robust assessment of past global temperature variations as model errors and biases may be averaged out (*Hegerl and Zwiers*, 2011). A measure of model pattern uncertainty has also been incorporated into some analyses (*Huntingford et al.*, 2006). Analyses using multi-model averages have detected well mixed greenhouse gas influences, but with other anthropogenic - non well mixed greenhouse gas - influences not being robustly detected (*Jones et al.*, 2013; *Gillett et al.*, 2013). Second, examining the net anthropogenic influence, rather than separating it into greenhouse gas and other anthropogenic influences, gives consistent results across a range of climate models with different analysis methodologies (*Gillett et al.*, 2013; *Ribes and Terray*, 2013). The attributed net trend of anthropogenic influences deduced from these studies (*Bindoff et al.*, 2013) was close to, or a little more than, the observed trend seen over the last 60 years. *Bindoff et al.* (2013) stated that results from using the net anthropogenic approach were 'much more robustly constrained', and used the optimal detection results to provide evidence to support the statement that it is 'extremely likely' (greater than 95% likelihood) that human activities caused more than half of the observed increase in near surface temperatures from 1951 to 2010. An earlier study, using a different but related detection methodology, also found that anthropogenic influences were robustly detected in observed temperature trends across several models, but the distinguishing of the greenhouse gas and sulphate aerosol components were not (*Hegerl et al.*, 2000). Nevertheless, the question remains why recent analyses, using ensembles of CMIP5 climate models, give such wide ranges of warming attributable to greenhouse gases and cooling attributable to other anthropogenic factors.

This study attempts to investigate why using different CMIP5 climate models in detection studies leads to such a variety of results. We use as our basis the main analysis of *Jones et al.* (2013), which applied an optimal detection analysis to variations of near surface temperatures over the period 1901 to 2010, using 8 CMIP5 models to deduce the contributions from changes in well mixed greenhouse gas concentrations, other anthropogenic influences and natural factors. Here we use a period ending in 2005 that enables the examination of a wider range of models than has been looked at before . The differences and similarities between the spatio-temporal response patterns of the models are explored in detail. We look at what choices in the basic methodology can be made to try and increase the robustness of the results.

This paper is structured as follows. The sources of observed and model data are described in the next section together with what model simulations were used. Section 3 describes the optimal detection methodology. The temperature responses and radiative forcings due to the different forcing factors are described in Section 4, noting any major differences which might lead to a diversity in the detection results. The results of the optimal detection analyses on the CMIP5 models are given in section 5. Section 6 describes the implications of the results on techniques to constrain transient climate response. There follows a discussion section and the conclusions.

## 2    Data

The HadCRUT4 (*Morice et al.*, 2012) data set of blended land air temperatures and sea surface temperatures are used for the observations. The dataset has a sophisticated error model, a component of which is an ensemble of realisations sampling one source of observational uncertainty. We use the median of this ensemble in this study. Including HadCRUT4's error model in an optimal detection analysis will be investigated in a future study.

Monthly mean near surface temperatures (TAS) from climate models were retrieved from the archive of the fifth phase of the Coupled Model Intercomparison Project (CMIP5) (*Taylor et al.*, 2012). Data from four CMIP5 experiments were obtained (Table 1). We obtained piControl data, to characterise climate internal variability, from 23 models that had at least 500 years of data available (Table 2) (Model details can be found in Table 9.A.1 in *Flato et al.*, 2013). We obtain data from the 15 models that also have historical, historicalGHG and historicalNat experiments that cover the 1900-2005 period (Table 2).

What forcing factors are included, and how they are implemented, differs across the models (*Collins et al.*, 2013; *Jones et al.*, 2013). Here all models included variations of the main well mixed greenhouse gases ($CO_2$, $CH_4$, $N_2O$, CFCs) in the historical and historicalGHG experiments. Variations in concentrations in tropospheric and stratospheric ozone are included in all the historical simulations, and some models' historicalGHG experiments (Table 2). The direct radiative effects (also known as aerosol-radiation interactions (*Myhre et al.*, 2013)) of $SO_4$ and carbonaceous aerosols are included in all the models. Indirect aerosol effects (also known as aerosol-cloud interactions (*Myhre et al.*, 2013)) are incorporated within most of the models but not all (Table 2). Similarly in some of the models, anthropogenic land use changes are not simulated. The radiative effects of changes in solar irradiance and stratospheric volcanic aerosols are included in all the models' historical and historicalNat experiments.

The experiments can be used to extract the climate change patterns associated with well mixed greenhouse gases, other anthropogenic factors and the influence of natural radiative forcing in an optimal detection analysis. For example, historicalOA (Table 1) responses are deduced by finding the difference between the ensemble means of the historical and the sum of the historicalGHG and historicalNat simulations. We can include more models than used in earlier studies (*Jones et al.*, 2013; *Gillett et al.*, 2013; *Ribes and Terray*, 2013) as we use data ending in 2005 rather than 2010. The models drawn from CMIP5 used here, a so called 'ensemble of opportunity', may not representatively sample the full range of possible model uncertainties (*Hegerl and Zwiers*, 2011), so there are limitations to the statistical interpretation of such an ensemble (*Knutti*, 2010; *Jones et al.*, 2013).

# 3   Optimal detection analysis methodology

We use multiple linear regression - in this case total least squares - of the simulated climate patterns $x_i$ against observed climate $y$ (Equation 1), optimized by projecting filtered patterns onto an estimate of the leading orthogonal modes of internal climate variability, which have been 'whitened' to normalise the noise characteristics and down-weight patterns with low signal to noise ratios (*Allen and Stott*, 2003).

$$y - v_0 = \sum_{i=1}^{I} \beta_i \left( x_i - v_i \right) \qquad (1)$$

The $\beta_i$ in Equation 1 are the regression scaling factors and $v_i$ and $v_0$ the internal variability components for the $i$th pattern and observations respectively. The details of the implementation are identical to that used in *Jones et al.* (2013), except for the periods considered, and specific models used.

Here the basis of orthogonal modes are the empirical orthogonal functions (EOFs) calculated from estimates of internal variability of the models, normally based on the model's own piControl. There are not long enough piControl experiments to produce EOFs for each model so we have used a common EOF basis approach, where the piControls from the CMIP5 models (Table 2) are combined to enable an estimate of a common EOF basis to be calculated (*Jones et al.*, 2013, and references therein).

A detection is deduced by testing the null-hypothesis that the scaling factor has a value of 0. If a pattern's regression scaling factor has an uncertainty range that does not cross 0, we conclude that the fingerprint pattern is detected (*Hegerl et al.*, 2007). A step towards attribution requires deducing if responses to the expected forcing factors are consistent with the observed change (*Mitchell et al.*, 2001). Previously, attribution consistency was examined by seeing if scaling factors have uncertainty ranges consistent with a value of 1, indicating that the responses from the models do not need to be significantly scaled up or down (*Hasselmann*, 1997; *Hegerl et al.*, 2007). More recently the constraint of scaling factors being consistent with 1 has been considered unnecessary for attribution, as long as any discrepancies are understandable within an expert judgement of the uncertainties (*Hegerl and Zwiers*, 2011; *Bindoff et al.*, 2013). An important stage towards attribution is to exclude other plausible factors as being alternative explainers of the changes (*Hasselmann*, 1997; *Mitchell et al.*, 2001). If many patterns are included in a multi-variate regression, there is an increased risk of an overfitting to the observations, which would bias the results. Patterns that are correlated with each other risk causing the scaling factors to be degenerate, with wide uncertainties. Using smaller numbers of patterns in the regression can lessen these problems. We should note that using smaller numbers of patterns makes stronger assumptions about equality of scalings on forcings since if the patterns used are composed of contributions from different forcing factors, then effectively the scaling factors for those components are assumed to be equal to each other. Using smaller numbers of patterns, and thus predictors, increases the risk of underfitting to the predictand. One conservative test for under/over fitting, and an important step for attribution, is to examine the residuals of the regression and test if the variability is consistent with estimates of internal variability (*Allen and Stott*, 2003).

# 4   Climate response patterns and radiative forcing

The model data analysed in this section are not masked by observational spatial coverage, as we are only concerned in an inter-model comparison at this stage. All the models show a gradual increase in historicalGHG global temperatures, with ensemble mean linear trends over the 1906-2005 period varying between 0.81 and 1.65K/century, with the rate of change at the end of the 20th century greater than the start of the century (Figure 1). The historicalNat simulations show a much smaller warming and then cooling over the twentieth century with overall trends over the whole period, ranging from -0.21 to 0.07K/century. The historical simulations all warm, with ensemble mean linear trends varying between 0.29 to 1.17K/century, but with more multi decadal variability than historicalGHG.

The ranges in the historicalOA trends over 1906-2005 are wider than the other experiments, varying between -1.25 to 0.01K/century (Figure 1). Some models (A, E, G, O) gradually cool to reach a minimum in temperatures around 1980 before then warming, others cool then stabilise after the 1980s (C, I, K, L, N) and some models continue decreasing in temperature to 2000 (D, F, H, J, M).

Differences in the models climate sensitivities and in the radiative forcing will contribute to the wide range in climate responses. One model index, the Transient Climate Response (TCR) - the global mean temperature change at doubling of $CO_2$ in a simulation with 1% per annum increase in $CO_2$ - is considered a useful model metric of climate sensitivity to compare multi-decadal scale changes in model response (*Flato et al.*, 2013). However, in practice the temperature response to different forcing factors will be more complex across models than the differences in TCR may suggest. There is no clear simple relationship between the models' TCR and the historicalGHG trend (Figure 2(a)) (*Gillett et al.*, 2013). One reason for this is that those models including ozone variations in the historicalGHG experiments (Table 2) may warm more, as ozone has a net warming radiative influence (*Myhre et al.*, 2013), which varies across the models (*Eyring et al.*, 2013). For instance models F and K warm more than other models with similar TCR (Figure 2(a)) but models C, L, M and N do not indicate an obvious warm bias. *Gillett et al.* (2013) came to a different conclusion, based on a different assessment of what CMIP5 models included ozone in their historicalGHG simulations, and suggested that the models varying responses to non $CO_2$ greenhouse gases will have contributed to the diversity in historicalGHG responses. There is no clear relationship between model TCR and historicalOA cooling (Figure 2(b)), reflecting a large spread in radiative forcing across the models, the range of model TCR, and in noise contamination due to the way historicalOA is derived.

There are large differences in the ERF (effective radiative forcing, adapted from *Forster et al.* (2013)) across the models (Figure 3), in particular with historicalOA having a wider range (-2.11 to 0.10Wm$^{-2}$) than historicalGHG (1.41 to 3.16Wm$^{-2}$) by the end of the 20th century. As for the temperature response, the historicalOA ERF will have some extra noise contamination due to the way it is derived.

The largest radiative influence that is not included in all models (Table 2) is due to indirect aerosol effects. Models that simulate the effect should tend to have a stronger cooling influence from aerosols than those models that do not. The historicalOA ERF for models that include indirect effects have a range of -2.11 to -0.58Wm$^{-2}$, while the models not including these aerosol processes have a smaller magnitude radiative influence by 2000 of -0.46 to 0.10Wm$^{-2}$. The models not incorporating aerosol indirect effects (A, B, G and O) have generally smaller magnitude historicalOA cooling trends over the 20th century (-0.40 to 0.01K/century) than the models that do include indirect effects (-1.25 to -0.38K/century). It has been previously noted that models which include indirect effects have historical simulations that appear to match the observed near surface temperature variations more closely than models that do not (*Jones et al.*, 2013; *Wilcox et al.*, 2013; *Ekman*, 2014).

Estimates of radiative forcing for ozone (based on a different methodology to that in *Forster et al.* (2013)) suggest a range of 0.17 to 0.44Wm$^{-2}$ for a small number of CMIP5 models (*Shindell et al.*, 2013). Another choice that is potentially important is whether or not land use changes are implemented (Table 2). Whilst not estimated by *Forster et al.* (2013), the IPCC assessment of land use radiative forcing for 2005 was $-0.15 \pm 0.10$ Wm$^{-2}$ (*Myhre et al.*, 2013). Those models not including land use variations in the historical simulations may warm more than those that do.

The implementation of aerosol physics differs substantially across the models (Table 12.1 in *Collins et al.*, 2013), from being prescribed in some models to varying degrees of interactivity in others. Despite the same source emissions dataset being used (*Lamarque et al.*, 2010) in most models, the differences in physics lead to the models having differently evolving concentrations of the aerosol species (*Wilcox et al.*, 2015). The aerosol optical depth (AOD) is a measure of the optical effects of sulphate aerosols as well as a number of other anthropogenic and natural aerosol species, and gives an indication of the evolution of the aerosols (*Wilcox et al.*, 2013; *Shindell et al.*, 2013; *Flato et al.*, 2013). The increase in global mean AOD from the 1940s to the end of the century (Figure 4(a)) is common between the models with some showing much stronger increases than others. Most models show a peak in AOD by the 1980s/90s but others increase to the end of the century. The differences between the models are more marked when comparing the contrast between a region covering North America, North Atlantic and Europe and the South East Asia region (Figure 4(b)). Some models (e.g. F) have a very strong change from the former to latter region whereas others (e.g. M) have a much smaller change.

All the historicalGHG responses show the typical higher latitude warming due to greenhouse gas increases, with most models showing an asymmetry of larger warming in the northern hemisphere (Figures 5 and 6). The historicalOA responses show somewhat more complex evolution, although the forced patterns of change will be more contaminated by noise because of the way historicalOA is derived. Some models show strong cooling in higher northern latitudes, but other models show more spatially uniform cooling which levels out near the end of the century.

Differences in the model responses can be demonstrated by looking at three models; B, D and F. Models B and D have the same TCR (Table 2), with similarities between the historicalGHG global mean responses (Figure 1), albeit with the spatial temporal pattern being somewhat more symmetric in D than in B (Figure 5). Model F has only slightly higher TCR than models B and D but has much more warming in the historicalGHG, especially over the high northern latitudes, reflecting a larger ERF (Figure 3) contributed to by the presence of ozone in Model F's historicalGHG. The historicalOA responses are very different for the three models. Model B shows little if any cooling globally or spatially, related to the lack of modelling of indirect aerosols. The D model has strong cooling that is again notably symmetric across the hemispheres, whilst model F has the strongest cooling globally, with most of the response across northern high latitudes.

One way to quantify the similarities and differences across the models is to look at the cross-correlations of the spatio-temporal data shown in Figures 5 and 6. The historicalGHG experiment has the most similar patterns between models with correlation coefficients varying between 0.68 and 0.97 (Figure 7). In contrast both the historicalNat and historicalOA experiments have intra model correlations that vary considerably, from -0.40 to 0.72 for historicalNat and from -0.35 to 0.87 for historicalOA, an indication of large differences between the response patterns. Noise from internal variability will mask some of the forced response, contributing to the pattern differences, particularly when the forced component of the response is weak, such as for historicalNat.

For the historicalGHG experiment all the models have high correlations ($> 0.8$) with the multi-model average, which are generally larger than the correlations with other models. This suggests that the multi-model average is representative of the response patterns of all the models. This is not the case for historicalNat where the biggest correlation of any of the individual models with the 'Mod.Avg' is 0.49. For historicalOA most of the individual models have higher correlations with the multi-model average than they have with the majority of the other models. For the two models where this is not the case (models B and G) the correlations with the 'Mod.Avg.' are very low, 0.15 and -0.19 respectively. Only five of the models have correlations with the 'Mod.Avg.' that are greater than 0.8, suggesting that the multi-model average is only representative of some of the models' other anthropogenic response patterns. Thus the multi-model average pattern for historicalOA may contain errors and biases due to the inclusion of some models with diverse and inconsistent response patterns.

# 5 Results of an optimal detection analysis

We examine large spatial and temporal scales over the 1906-2005 period, using the same standard spatio-temporal filtering as described in *Jones et al.* (2013) and the optimal detection methodology described in Section 3. We project all the model data onto the same spatial grid as HadCRUT4, calculate non-overlapping 10 year means, masking the model data to have the same spatial coverage as the observations and project it onto spherical harmonics (T4) to filter out smaller than 5000km spatial scales (*Stott and Tett*, 1998).

Combining 250 years from each of the 23 models' piControls (Table 2) creates a common EOF basis with 69 degrees of freedom. Separate 250 years from each model's piControl are used for the uncertainty analysis and residual consistency testing. The scaling factors can be sensitive to the choice of truncation of the EOF space. Including higher order EOFs will allow a higher proportion of the variance explained of the original data to be retained. This will be at the cost of including variability modes that the required forced response patterns do not project strongly upon, thus adding noise but not any signal, i.e., reducing the signal to noise ratio (*Hasselmann*, 1997). As such a balance needs to be struck between maintaining a strong signal and retaining patterns that are relatable to the original data (*Tett et al.*, 2002). An EOF truncation of 40 retains 97% of the variance of HadCRUT4. Using 40 EOFs also captures 97 to 99%, 96 to 98%, 84 to 97% and 78 to 93% of the variance of the historicalGHG, historical, historicalNat and piControl simulations, respectively. With a high proportion of the variability of the patterns captured by the first 40 EOFs, this seems a reasonable truncation choice for the following analysis, but the sensitivity to the choice is investigated (Supporting information, Figure S1).

The first analysis we look at is the three way regression that uses the historical, historicalGHG and historicalNat simulations to produce the scaling factors; $\beta_{historical}$, $\beta_{historicalGHG}$ and $\beta_{historicalNat}$ (Equation 1). These are then transformed to the required scaling factors for well mixed greenhouse gas forcings ($\beta_G$), other anthropogenic forcings ($\beta_{OA}$) and natural forcings ($\beta_N$), following equation 2 in *Jones et al.* (2013) (Also see Supporting Information). This approach has been used on CMIP5 data previously (*Gillett et al.*, 2013; *Jones et al.*, 2013; *Ribes and Terray*, 2013). Similar studies in the past have used different experiments to deduce the G, OA and N scaling factors (e.g., *Allen et al.*, 2000; *Tett et al.*, 2002; *Stott et al.*, 2006)

Of the 15 models examined there are only 8 cases where G is detected, with $\beta_G > 0$ (Figure 8(a)), most of them across a wide range of EOF truncations (Supporting information, Figure S1). In the cases where G is detected a test of the residual variability being consistent with an estimate of the internal variability (*Allen and Tett*, 1999; *Allen and Stott*, 2003) is passed at the two sided 10% level. Only 5 of the models (C, I, J, L and O) have G scaling factors consistent with 1, suggesting those models do not need significant scaling up or down.

Of the 8 analyses on individual models that detect G, other anthropogenic (OA) factors are detected in 5 (using models D, F, J, L and O), none with values consistent with 1, although this is somewhat sensitive to EOF truncation choice (Figure S1). Natural factors (N) are detected in 7 of the cases that detect G, with 5 having values consistent with 1. Only for the analyses using models F, J, L and O are all three patterns detected simultaneously. The analysis using the multi-model average - calculated as the average of the models' ensemble means ('Mod.Avg.' in Figure 8) - G and N are detected, both with scaling factors consistent with 1, but OA is not detected. This is a robust result across choice of EOF truncations. It should be noted that the multi-model average analysis does not include a measure of model pattern uncertainty (e.g., *Huntingford et al.*, 2006).

The patterns and scaling factors can be used to reconstruct the scaled global mean temperatures of the different forcing factor responses, i.e., $\beta_i(x_i - v_i)$ (*Allen and Stott*, 2003). Where G is detected there are a varied range of scaled trends for the 1906-2005 period (Figure 8(b))), from 0.36 to 0.82 K/century (model E) to 1.32 to 4.22 K/century (model D). Confidence in the observed changes, with a trend of 0.65K/century, being mostly attributable to G will be strengthened, for an individual model, if the scaling factors are consistent with 1 (Figure 8(a)).

For the model analyses that also detect it, OA has also wide linear trend ranges from -3.69 to -0.71 K/century (model D) to -0.48 to -0.09 K/century (model J). Where N is detected the trends are very close to 0K/century with the largest magnitude cooling trend around -0.33 to -0.09K/century (model E). The analysis of 'Mod.Avg.' produces scaled trends for G of 0.60 to 1.16 K/century, for OA of -0.45 to 0.12 K/century and for N of -0.07 to -0.02 K/century.

These results are generally in line with the analysis in *Jones et al.* (2013), which examined the 1901-2010 period with a smaller number of CMIP5 models. What differences there are between the studies appear to be due to the change in choice in period and choice of EOF truncation. Using the 1901-2010 period, for these models, gives very similar results to *Jones et al.* (2013), even though the common EOF basis is constructed from a different sample of CMIP5 models (Supporting information, Figure S2).

These results can also be compared with those of *Gillett et al.* (2013) and *Ribes and Terray* (2013) which also applied an optimal detection methodology to CMIP5 models and HadCRUT4 to deduce the scaling factors for G, OA and N. While both studies used the same spatio-temporal filtering as used here, there are differences in the analyses. For instance, different sets of the CMIP5 models were used, *Gillett et al.* (2013) used the period 1861-2010 for their main analysis and *Ribes and Terray* (2013) applied an alternative optimal fingerprinting method in addition to the more standard approach. These differences, as well as other methodological choices such as choice of EOF truncation, mean the results will differ. However, the results shown in Figure 8, support these studies in showing little consistency in the magnitude of the scaled greenhouse gas warming across a sample of CMIP5 models. Using the multi-model average was considered to be the most robust result (*Jones et al.*, 2013; *Gillett et al.*, 2013), but it is then legitimate to question the confidence of the magnitude of the attributed greenhouse gas warming when another important forcing factor with known strong radiative effects is not detected at the same time. As other anthropogenic influences are not robustly detected, is the factor not important for 20th century temperature changes, are there errors or biases in the other anthropogenic response patterns, are other important factors not being included, or is the detection analysis methodology flawed?

One of the expectations of optimal detection techniques is that they should bring the magnitude of the model responses into closer agreement, as long as there are no major differences in the spatial patterns and temporal evolution of the models' responses (e.g., *Hegerl et al.*, 2007). However, as also seen in recent studies using CMIP5 models (*Jones et al.*, 2013; *Gillett et al.*, 2013; *Ribes and Terray*, 2013), the range of G and OA trends after scaling (Figure 9(b)) is larger than before scaling (Figure 9(a)).

One issue that has been often overlooked in recent detection and other related regression studies is a statistical examination of the signals being used as predictor variables to deduce if their inclusion is optimum. The historical (GOAN) and G patterns for most models are strongly correlated with each other and with the HadCRUT4 pattern with values typically around 0.8 (Table 3). With very strong correlations between GOAN and G there is the danger of degeneracy within a regression and large (compensating) uncertainties in the scaling factors for the patterns concerned (*Allen and Tett*, 1999). The N signal, in contrast, is only weakly correlated with GOAN and G, reducing the risk of degeneracy. However, N is also weakly correlated with HadCRUT4, which suggests it has lower importance as an explanatory variable.

Predictor patterns with weak magnitudes relative to the noise present can lead to biased estimates - when using ordinary least squares (OLS) - or unbiased but very uncertain results - when using total least squares (TLS) as used here (*Allen and Stott*, 2003). The relative strengths of underlying signals relative to the internal variability can be measured by looking at the signal to noise ratios (SNR) (see Table 4 - calculated following the technique in *Tett et al.* (2002)). All models have GOAN and G patterns with SNRs over 2, many considerably larger. In contrast, no model has a N signal with SNR above 2 and seven have a SNR less than 1.5. Of those seven models, their optimal detection analyses produce poorly constrained scaling factors for all 3 signals in 5 of the cases (Figure 8(a)). This illustrates the consequence of including low SNR patterns in a multivariate regression (*Allen and Stott*, 2003) which can increase uncertainties in the scaling factors. It also suggests that for some purposes, it would not be appropriate to include N as a separate predictor pattern due to its generally weak SNR.

In view of these factors, it is usual in multivariate analyses to consider if one or more of the variables could be discarded from the regression. That the emergent responses to forcing factors from climate models are used, should provide further confidence in the results and is a major advantage of this approach over other regression approaches to attributing causes to past climate changes (*Bindoff et al.*, 2013). So, rejecting the inclusion of a pattern, when there are good physical reasons for it to be included in an analysis, could be considered unreasonable (*Allen et al.*, 2006). Thus excluding a factor's climate response from the analysis should not be done lightly.

One set of techniques (*Tett et al.*, 1999) that has previously been regularly used in optimal detection climate studies is based on degeneracy tests (*Mardia et al.*, 1979). Several tests examine the component patterns within the predictor variables and allow some measure of how many patterns should be allowed in the regression. The aim is to not have too many signals which can increase the chance of overfitting in the regression. We use principal component analysis tests (pages 243-245 in *Mardia et al.*, 1979), implemented as in *Tett et al.* (1999), that examine the independence of the predictor variables to deduce which are of less importance and thus could potentially be discarded from the regression. None support using more than two signals for any of the models. Hence we consider all combinations of two or less predictands that can be constructed from the available, historical, historicalGHG and historicalNat experiments. This enables the examination of alternative plausible causal climate factors. If any of these are ruled out the confidence in the attribution of a dominant greenhouse gas influence is strengthened (*Hasselmann*, 1997).

## 5.1 Regression with GOA and N

This combination is based on the two way regression using historical and historicalNat to deduce anthropogenic (GOA) and natural (N) scaling factors (Figure 10).

How the signals are transformed and how sensitive the results are to choice of EOF truncation are summarised in the supporting information and Figure S3. The derived anthropogenic spatial temporal responses (historicalGOA) are generally quite similar across the models (Supporting Information Figure S9), apart from models M and G which have fairly low spatio-temporal correlations with other models. GOA, is detected by all the model analyses, although in 3 of the cases (using models F, J and M) the residual consistency test fails, suggesting possible under/over fitting to the observations. Only when using 3 models is $\beta_{GOA}$ found to be consistent with 1. N is detected in nine of the model analyses. This signal combination was also examined by *Gillett et al.* (2013) and *Ribes and Terray* (2013) and was found to give more robust results than the G,OA,N combination. The GOA signal was detected in the analysis of all 9 models considered by *Gillett et al.* (2013) and in the analyses of 6 out of the 10 models considered by *Ribes and Terray* (2013). *Ribes and Terray* (2013) also analysed global ten year means as a climate index and found analyses of nine out of 10 models detect GOA (as reported by *Bindoff et al.*, 2013).

The scaled temperature trends for GOA are tightly constrained, being near to the observed trend, varying from 0.40 to 0.59K/century (model L) to 0.64 to 0.77K/century (model G). The 'Mod.Avg.' analysis detects both GOA and N with values consistent with 1 and with scaled trends for GOA of 0.57 to 0.73K/century and N of -0.05 to -0.01K/century for the 1906-2005 period. The scaling factors for the individual models and 'Mod.Avg.' are largely in line with what was found by *Gillett et al.* (2013) and *Ribes and Terray* (2013) despite the differences between the analyses. The reconstructed trends from the *Gillett et al.* (2013); *Ribes and Terray* (2013) studies (Figure 10.4 in *Bindoff et al.*, 2013), also found a fairly consistent attributed trend due to total anthropogenic influences to be near the observed warming trend, albeit for the different period of 1951-2010.

On this evidence the two way regression of anthropogenic and natural, GOA and N, seems to be a more robust technique for attributing past climate. However, there are two issues to be considered. First, while the scaled GOA trends of the CMIP5 models are in close agreement, not all of the models have scaling factors consistent with 1 or pass the residual consistency test. This may reduce the confidence in the attribution of the observed changes to anthropogenic influences for those models. Secondly, the models will have quite different contributions from G and OA to produce the same scaled GOA trends. To demonstrate this second point we can estimate individual contributions to the scaled GOA trends from G and OA with a simplifying assumption. The reconstructed scaled anthropogenic temperatures, can be deduced from Equation 1 as $\beta_{GOA}(x_{GOA} - v_{GOA})$. If we assume that $v_{GOA}$ does not contribute much to the trend of GOA, this can be expanded to $\beta_{GOA}x_G + \beta_{GOA}x_{OA}$, enabling an estimate of the contributions from G and OA. Figure 11 shows the estimated G,OA and N contributions to the scaled trends deduced from the G,OA,N and the GOA,N analyses. Inspection of Figures 8(b) and 11(b) demonstrates that this approximation of the G,OA and N trends is reasonable. The contributions from G and OA to the scaled GOA (Figure 11(b)) show that even when the scaled net anthropogenic warming is fairly consistent across the models (Figure 10(b)), the G and OA contributions are far from consistent. The estimated contribution from G ($\beta_{GOA}x_G$) across the models ranges from 0.63 to 0.83K/century (model B) to 1.84 to 3.04K/century (model F). The 'Mod.Avg.' result, for the GOA, N analysis, suggests a larger magnitude G warming and OA cooling than the G, OA, N analysis does.

The net anthropogenic warming result raises some pertinent questions. Because of the range in the magnitude of the responses, it will often not be possible to have self consistency ($\beta \approx 1$) for all the models at the same time as a similarity of scaled trends between the models. Thus, as a step towards formal attribution, it may not be a requirement that scaling factors are consistent with 1 for all the model analyses at the same time. This is consistent with the views that $\beta \approx 1$ should not be a strong constraint for attribution (*Hegerl and Zwiers*, 2011; *Bindoff et al.*, 2013) and that agreement between models' scaled trends is important for robust attribution (*Hegerl et al.*, 2007; *Bindoff et al.*, 2013). However, given the varied G and OA contributions to the consistent scaled GOA trends across the models (Figure 10(b)), together with the limited number of models with scaling factors consistent with 1 (Figure 11(a)), it must be of concern whether the agreement is an artefact or not (Section 6.1.2 in *Allen et al.*, 2006).

## 5.2   Regression with GOAN

When using only the historical experiment, the single pattern of GOAN (Supporting information, Figure S4) is detected in each of the 15 model analyses, with the scaled trends generally close to the observed trend. Several of the analyses produce lower scaled trends and fail the residual consistency test. This indicates that some model's GOAN patterns are unable to be matched with the observed pattern by scaling alone.

## 5.3   Regression with G

When using only the historicalGHG experiments, the single pattern of G (Supporting information, Figure S5) is detected in each of the 15 model analyses, with scaling factors almost always below 1. This is not surprising as the scaled trends are very close to the observed warming, and there is no factor to partially offset the greater G warming (*Allen et al.*, 2006).

## 5.4   Regression with N

The single pattern of N, derived from historicalNat alone (Supporting information, Figure S6), is either not detected or fails the residual consistency test across all the model analyses. This is important in the attribution of anthropogenic influences as it means natural external factors alone cannot explain the observed changes.

## 5.5   Regression with G and N

Using the experiments historicalGHG and historicalNat in a two way regression both G and N (Supporting information, Figure S7) are robustly detected except for the analyses of a few models, with G having low scaling factors as N is unable to offset much of the G warming. When different combinations of patterns are available in a regression of a physical process it is simplest to choose those that incorporate all the known major forcing factors (*Allen et al.*, 2006).

## 5.6   Regression with G and OAN

The analysis of G,OAN uses the regression of the historicalGHG and historical experiments to transform the scaling factors for G and OAN (Supporting information). This combination has previously been investigated, albeit for precipitation changes, in *Wu et al.* (2013). In each of the 15 model analyses G is detected, with 10 of the cases having G scaling factors consistent with 1 (Figure 12(a)). OAN is detected in all but one of the analyses, but is only consistent with a scaling factor of 1 in five of the cases. G and OAN are robustly detected across EOF truncations for each of the 15 models, with 10 of the cases detecting G across all truncations (Supporting information, Figure S8). The scaled G temperature trends (Figure 12(b)) are much more in agreement than in the G,OA,N analysis. There is still a substantial range, however, with the trend of G varying between 0.57 to 0.78K/century (model G) to 1.12 to 1.67K/century (model N). Similarly OAN has trends closer together than OA were in the G,OA,N analysis with ranges varying between -0.10 to 0.08K/century (model G) to -1.00 to -0.49K/century (model N). For the analysis using the CMIP5 model mean, 'Mod.Avg.', both G and OAN are detected with scaling factors consistent with 1, with attributed trends of 0.87 to 1.22K/century for G and -0.54 to -0.22K/century for OAN.

An examination of the cross-correlations of the spatio-temporal patterns gives a range of -0.11 to 0.92, but with most models appearing to have similar historicalOAN patterns (Supporting Information Figure S9), apart from models B,G and O. The marginal closer agreement between the historicalOAN patterns than the historicalOA patterns is partially due to the higher SNR of the former patterns (Table 4). Whilst the trend of scaled G (Figure 13(b)) is much more constrained than in the G,OA,N analysis, it still has about the same range as the unscaled G trend (Figure 13(a)). However the OAN scaled trend is in closer agreement than the unscaled OAN trend.

The scaling factors for G and OAN can be used to estimate the contributions from G ($\beta_G x_G$), OA ($\beta_{OAN} x_{OA}$) and N ($\beta_{OAN} x_N$) to the scaled trends (Figure 11(c)). The G,OAN analysis has the lowest spread of contributions from G and OA for the three analyses shown. This could suggest that this combination may be a more robust technique than either the three way regression (G,OA,N) or two way regression (GOA, N). However, close consistency of scaling factors does not necessarily mean the results are more accurate, as there may be consistent biases in the methods. For instance any differences in the uncertainty in the magnitude of forcing and response to natural and other anthropogenic factors in the G,OAN analysis cannot be accounted for.

As there is only one reality to test the models against, which technique is more robust is not easy to check for. One option is to do perfect model tests, where the patterns being investigated are known before time, for instance using climate models as surrogates for the observations (For example, *Stott et al.*, 2003; *Ribes and Terray*, 2013). One can then examine how well the models attribute the forcing components in other models.

### 5.7 Perfect model results

From the 15 CMIP5 model used in this study there are 72 historical simulations that can be used as predictands or surrogate observations. For each of the 15 models being used as predictors or 'detector' models the analysis is repeated on each of the surrogate observations - not including the historical simulations from the same model. Then in each case the scaled temperature trends can be compared to the surrogate model's own estimate of that forcing contribution. The fraction of detections and scaled trends consistent with the expected trend (when the residual test for consistency is also passed) is calculated for each detector model (Figure 14(a,b)) as well as for each surrogate observation model (Figure 14(c,d)). Thus it is possible to examine how faithful each model is in attributing influences in the other models, and in how well each model's own forced components are attributable by the other models.

For the G,OA,N combination (Figure 14(a)) G is detected in the surrogate models more than 80% of the time for only 3 of the models, and only once is OA detected in the surrogate models more than 80% of the time. In contrast G and OAN are detected when using 11 and 10, respectively, of the models more than 80% of the time in the analyses of the G,OAN combination (Figure 14(b)). Several models are very poor, < 10%, at providing patterns that can detect G in the G,OA,N analysis (Models B, D, H and M). In contrast in none of the model analyses is G detected less than 60% of the time for the G,OAN combination. Based on this analysis the G,OAN combination is more discriminating than using the G,OA,N combination.

The fraction of scaled trends consistent with the expected trends are generally higher in the G,OAN analysis than the G,OA,N analysis, but there are a number of models where that is not the case. For instance, for the G,OA,N combination, when 6 of the detector models are used there are higher G consistency fractions than when the same models are used for the G,OAN analysis. From the view point of the models acting as surrogate observations (Figure 14(c,d)), for any given model acting as the predictand, there is a higher fraction of the predictors that make detections and, except for one predictand model, a higher fraction of predictor trends consistent with the expected trends in the G,OAN analysis than in the G,OA,N analysis.

In other words some models are better at being detectors than others and some models better able to have their component signals detected than others. Overall the G,OAN analysis appears to be more skilful in detecting and estimating the true G trend in the surrogate observations, than the G,OA,N analysis. As noted for the analysis of G,OA,N against the observations, the analyses on models with low SNR, <1.5, for historicalNat (Table 4), tend to detect G with lower frequencies than analyses on models with higher SNR.

## 6 Implications for observationally constrained TCR

Optimal detection analysis results have been used to attempt to provide observationally constrained estimates of future warming (often called the 'ASK' approach after *Allen et al.*, 2000; *Stott and Kettleborough*, 2002; *Kettleborough et al.*, 2007). This applies scaling factors for G to future warming trends, by assuming that a model that is over/under responding in the past will do the same in the future. This technique can also be applied to the Transient Climate Response (*Stott and Forest*, 2007), i.e. $\beta_G$TCR, and has been used to provide estimates of observationally constrained TCR (*Stott and Jones*, 2012; *Gillett et al.*, 2012; *Stott et al.*, 2013; *Gillett et al.*, 2013). *Gillett et al.* (2013) estimated the optimal detection constrained TCR, based on the multi-model mean analysis, to be 0.9 to 2.3K (5-95%), which contributed to the IPCC's assessment that TCR 'is likely in the range 1°C to 2.5°C' (*Bindoff et al.*, 2013; *Collins et al.*, 2013).

As expected the wide, often unconstrained, ranges of $\beta_G$, in the G,OA,N analysis, produce wide ranges of scaled TCR (Figure 15(a)). Where G is detected the scaled TCR has ranges that vary between 0.72 to 1.65K (model E) to 2.53 to 8.12K (model D). The lack of consistency across models is similar to what was found by *Gillett et al.* (2013). The constraining of future warming based on the analysis on a single climate model could be misleading if the impact of model diversity is not acknowledged.

The regression analysis used by *Bindoff et al.* (2013) to support the anthropogenic warming attribution statement was the two way case of GOA, N. In principal the scaling factors for GOA, from a GOA, N analysis (*Gillett et al.*, 2013), could also have been used to constrain the TCR if one assumes that $\beta_{GOA} = \beta_G$. The ranges of scaled TCR ($\beta_{GOA}$TCR) are considerably varied (Figure 15(b)) across the models, from 1.09 to 1.65K (model G) to 3.17 to 5.03K (model D).

The ranges of scaled TCR ($\beta_G$TCR) deduced from the G,OAN analysis (Figure 15(c)) are less spread than in the two previous analyses, from 0.98 to 1.44K (model G) to 1.91 to 2.75K (model C). There is, however, still limited consistency across the models and it is arguable that TCR after scaling is not better constrained than before scaling. The scaled multi-model average of TCR for the G, OA, N analysis (Figure 15(d)), 1.07 to 2.06K, is similar to that reported in *Gillett et al.* (2013). Some of the uncertainties associated with the historicalGHG trends not having simple relationships with TCR (Figure 2) were included by *Gillett et al.* (2013) but not by us, so the spread in scaled TCR here is slightly smaller. The scaled multi-model average TCR is higher for the GOA,N and G,OAN analyses, 1.84 to 2.40K and 1.54 to 2.17K respectively.

The wide variety of scaled TCR values across models and different choices in the analysis has been seen previously (*Gillett et al.*, 2012; *Stott and Jones*, 2012; *Gillett et al.*, 2013). The results presented here strongly imply that considerable uncertainties exist with the specific use of a three way regression which includes N as a separate pattern. This highlights the importance of not putting too much emphasis on one result when there are sensitivities to analysis choice and to the uncertainty in the climate response to given forcing factors.

# 7 Discussion

The method of optimal detection relies on a number of assumptions. The most important is that patterns can be linearly combined (*Gillett et al.*, 2004) and that while response magnitudes are uncertain there are no errors in the spatio-temporal patterns. The examination of response patterns in Section 4 clearly demonstrates that there are a wide variety of response spatially and temporally across the models, especially for non well mixed greenhouse gas anthropogenic factors. The limited consistency in scaled temperature trends across the models in the observational analysis and in the perfect model tests is of concern. As some of the model responses are inconsistent with each other, the historicalOA multi-model average is not as representative of the models as the historicalGHG multi-model average, and as such its use will not give as robust results as once was considered (*Gillett et al.*, 2013; *Jones et al.*, 2013).

Clearly further work is required to address the issues raised in this paper; the importance of experimental design and analysis methodology; and how to incorporate pattern uncertainty. Other approaches related to optimal detection may be helpful to respond to these concerns. One technique that has been used in multi-model approaches is EIV, Error in variables, (*Huntingford et al.*, 2006). The method outlined in *Huntingford et al.* (2006) uses intra model variability to estimate the uncertainties in the shape, but not the magnitude, of the spatio-temporal patterns which can be included as an extra component in the uncertainty analysis. A related approach, that it is claimed to be an improvement on the *Huntingford et al.* (2006) technique and that can incorporate different sources of uncertainty, may be a promising development (*Hannart et al.*, 2014). However, techniques that try to include measures of model uncertainty often rely on sampling an 'ensemble of opportunity' (e.g., CMIP5), which may not be representative of the population of all physically plausible models (*Hegerl and Zwiers*, 2011). Another related approach to optimal detection described here is the recently proposed regularised optimal fingerprinting (ROF) technique (*Ribes et al.*, 2013). The technique derives a regularized estimate of the covariance matrix that is more accurate than other methods (*Bindoff et al.*, 2013), with a "trade-off between bias and variance" (*Ledoit and Wolf*, 2004). Techniques which attempt to account for model/observational discrepancies (*Harris et al.*, 2006), apply Bayesian methods (*Lee et al.*, 2005), or screen models depending on a measure of their quality (*Santer et al.*, 2012) may be helpful.

The optimal detection analysis applied in this study uses a common EOF basis, as the different CMIP5 models did not have enough available piControl data to characterise each model's internal variability well enough. As the CMIP5 models sample a wide range of interannual to interdecadal temperature variations due to internal variability (Figure 5 in *Jones et al.*, 2013), using a common EOF basis will be sub-optimal for analyses on individual models. A preferred approach, for a regression analysis with an individual model, would be to be able to characterise the internal variability of that model, which could be done with much longer length piControls (*Jones et al.*, 2013). The techniques to investigate if inclusion of predictor signals is superfluous or not should be considered, not only for optimal detection studies, but also for alternative regression analyses that have also been used to estimate contributions to past climate (e.g., Section 10.3.1.1.3 in *Bindoff et al.*, 2013), to avoid misinterpretation of results.

A set of model experiments has been recommended by *Ribes et al.* (2015) for inclusion in the next phase of the Climate Model Intercomparison Project, CMIP6 (*Meehl et al.*, 2014). The proposed experiments are different to what have been used in previous optimal detection analyses; anthropogenic and natural forcings together, natural only and aerosol only forcing factors. *Ribes et al.* (2015) state that these would enable the deduction of contributions from greenhouse gases, aerosol and natural forcing factors. By better characterising the aerosol climate response, it is claimed, the 'greenhouse gas' attributable warming would be better constrained. *Ribes et al.* (2015) recommended that large numbers of ensemble members of historicalNat would be preferable in future analyses. They concluded that, given a total number of 25 simulations to apportion to the different historical forcing experiments in their perfect model analysis, results were more reliable with order 10 historicalNat ensemble members. However, this is much higher than were provided by any institution for CMIP5 (Table 2). It is doubtful that many institutions would have the resources for this recommendation, especially given the general increase in the number of experiments that could be done for CMIP6 (*Meehl et al.*, 2014). In their perfect model analysis, *Ribes et al.* (2015) were limited to using models with strong aerosol cooling effects. They explain that the optimum number of ensemble members could have been different if models with different magnitude aerosol cooling were considered. *Ribes et al.* (2015) also did not examine the impact of the detector model having different patterns than the "truth" model. It is thus unclear how appropriate the recommendations are in practice.

To encourage as many models to be included in CMIP6 for detection studies as possible, more practical recommendations of numbers of experiments and of ensemble members should be considered, such as focusing on initial condition ensembles of historicalGHG and historicalOAN. Experimental design should take into account the likely strength of the response patterns. If a very large number of ensemble members would be needed to reasonably characterise the weak response of some forcing factors, it should be re-considered if the experiment is a high priority or not. To also help better characterise the responses to forcing factors of particular interest, experiments should be designed to limit having to derive response patterns from other experiments. The impact this may have on the total number of simulations that would be produced will need to be accounted for. Planning multi-model experimental designs (such as CMIP6 (*Meehl et al.*, 2014)) will also require considering the importance of a range of variables, on different time and spatial scales, that may have responses with differing SNR. For instance the historical global precipitation response to natural forcing factors is much stronger than that to well mixed greenhouse gases (*Bindoff et al.*, 2013) and analyses with different choices of spatial/temporal filtering may produce patterns with much stronger SNR.

Why earlier studies (such as reported in *Hegerl et al.*, 2007) gave more consistent scaled trends for different models, when deducing the greenhouse gases, other anthropogenic and natural factors contributions to 20th century observed warming, is still an open question. Before CMIP5, total solar irradiance datasets were used that had larger increases over the twentieth century, which would have lead to stronger natural response patterns. There were simpler experimental designs that improved the SNR of the other anthropogenic factors, and the model response patterns may have been less varied than in CMIP6.

# 8 Conclusions

A wider range of climate models have been included in an optimal detection analysis of large spatial and temporal scale variations of 20th century surface temperature variations than ever before. Only the analyses of eight of the fifteen CMIP5 models examined detect a separate greenhouse gas signal and of those only 5 also detect the influence of other anthropogenic factors in a standard analysis design. The scaled greenhouse gas trends show a wide range across the models with little consistency, supporting previous studies. This variety of results appear to be largely down to two factors; 1) The inclusion of patterns with low signal to noise introduces noise into the regression and 2) differences in the spatio-temporal patterns across the models which are irreconcilable in the optimal detection analysis as it is currently designed. In particular the temporal and spatial patterns of response to the non greenhouse gas anthropogenic factors are more varied across the CMIP5 models than the response to greenhouse gases. Using an alternative analysis design, by using historical and historicalGHG experiments, we find that well mixed greenhouse gases are detected with all models and other anthropogenic and natural influences combined in the vast majority of models. Of the observed warming of 0.65K per century, the multi model average analysis (not including a measure of model pattern uncertainty (*Huntingford et al.*, 2006; *Hannart et al.*, 2014)) attributes a well mixed greenhouse gas warming of 0.86 to 1.22K per century and other anthropogenic and natural cooling of -0.54 to -0.21K per century. However with the models included in the fifth phase of the climate model intercomparison project sampling a wider range of physical and chemical processes than ever before, the more distinct responses of the models still make it difficult to get consistent attributed trends.

There are a number of ways in which the utility of future model intercomparison projects could be improved for attribution studies of large scale temporal and spatial variations of near surface temperatures. The first would be to ensure as much model data is available as is possible. i.e., to have long - thousands of years - piControl simulations, to better constrain the multi-decadal variability from each model. And to have large initial condition ensembles to better characterise the forced responses. Second, if for most models very large initial condition ensembles are not possible, to consider designing experiments that avoid predictors with low signal to noise. Third, to have a set of experiments which do not require the differencing of two or more experiments to deduce the response to wanted forcing factors. This will reduce uncertainties in the responses, for instance helping to distinguish the other anthropogenic or aerosol response patterns from greenhouse gases. However, depending on the specifics of the experimental design, this may have a cost with regards to the total number of simulations to be produced. Fourth, to have as consistent application as possible, across the models, of what forcing factors are included in each experiment, such as indirect aerosols being applied by all models. This will aid the interpretation of the experiments responses, although it may be technically challenging from some Earth system models. Finally, a thorough quantifying of the radiative forcing from the major individual forcing factors contributing to the experiments for each model would greatly contribute to the understanding of model response patterns (*Forster et al.*, 2013; *Andrews*, 2014). Of course in the planning of future model intercomparison projects, consideration should also be given to the requirements of other analyses which examine other variables on different time and space scales than examined in this study.

Given the wide range of model responses to not only greenhouse gases but also other anthropogenic factors, consistency in the attribution of trends, across models, is unlikely to happen - with the current formulation of the analysis. There is a challenge to the detection and attribution community to improve how pattern spatio-temporal uncertainty is dealt with in the standard optimal detection methodology, and how to interpret the variety of results when using an 'ensemble of opportunity' that may poorly sample model diversity (*Hegerl and Zwiers*, 2011).. Approaches that try to account for pattern uncertainties (*Huntingford et al.*, 2006; *Hannart et al.*, 2014) should be examined further. The results presented here do not throw into question that well mixed greenhouse gases are the dominant influence on changes in near surface temperatures over the last 100 years or so. But a return to more thorough use of techniques in regression studies and a better understanding of differences between the models will help to constrain what has happened in the past and give more confidence in climate projections.

## acknowledgments

We acknowledge the Program for Climate Model Diagnosis and Intercomparison and the World Climate Research Programme's Working Group on Coupled Modelling, which is responsible for CMIP, and we thank the climate modeling groups for producing and making available their model output. We thank Ben Booth, Laura Wilcox and Annica Ekman for useful discussions about the CMIP5 aerosol modelling. We thank Tim Andrews for providing CMIP5 TCR data and Piers Forster for ERF data. We are grateful to Natalie Mahowald for providing aerosol optical depth data for the CCSM4 model. We wish to thank the reviewers of this manuscript for their useful comments. The CMIP5 data used in this study were obtained from `http://cmip-pcmdi.llnl.gov/cmip5/` and were up to date as of March 2013. All models used were 'p1' physics versions. Version numbers of the data retrieved are available on request from the lead author (gareth.s.jones@metoffice.gov.uk). HadCRUT4 data was obtained from `http://www.metoffice.gov.uk/hadobs`, version 4.1.1.0, retrieved March 2013. The work of the authors was supported by the Joint UK DECC/Defra Met Office Hadley Centre Climate Programme (GA01101)

## References

Allen, M. R., and P. A. Stott (2003), Estimating signal amplitudes in optimal fingerprinting, part I: Theory, *Clim. Dynam.*, *21*(5-6), 477–491.

Allen, M. R., and S. F. B. Tett (1999), Checking for model consistency in optimal fingerprinting, *Clim. Dynam.*, *15*, 419–434.

Allen, M. R., et al. (2006), Quantifying anthropogenic influence on recent near-surface temperature change, *Surv. Geophys.*, *27*(5), 491–544.

Allen, M. R., P. A. Stott, J. F. B. Mitchell, R. Schnur, and T. L. Delworth (2000), Quantifying the uncertainty in forecasts of anthropogenic climate change, *Nature*, *407*(6804), 617–620.

Andrews, T. (2014), Using an AGCM to Diagnose Historical Effective Radiative Forcing and Mechanisms of Recent Decadal Climate Change, *J. Climate*, *27*(3), 1193–1209.

Bindoff, N. L., et al. (2013), Detection and Attribution of Climate Change: from Global to Regional., in *Climate Change 2013:The Physical Science Basis. Contribution of Working Group I to the Fifth Assessment Report of the Intergovernmental Panel on Climate Change*, edited by T. F. Stocker, et al., pp. 867–952, Cambridge University Press, Cambridge, U.K.

Collins, M., et al. (2013), Long-term Climate Change: Projections, Commitments and Irreversibility, in *Climate Change 2013:The Physical Science Basis. Contribution of Working Group I to the Fifth Assessment Report of the Intergovernmental Panel on Climate Change*, edited by T. F. Stocker, et al. , pp. 659–740, Cambridge University Press, Cambridge, U.K.

Ekman, A. M. L. (2014), Do sophisticated parameterizations of aerosol-cloud interactions in CMIP5 models improve the representation of recent observed temperature trends?, *J. Geophys. Res.*, *119*(2), 817–832, doi:10.1002/2013JD020511.

Eyring, V., et al. (2013), Long-term ozone changes and associated climate impacts in CMIP5 simulations, *J. Geophys. Res.*, *118*, 5029–5060, doi:10.1002/jgrd.50316.

Flato, G., et al. (2013), Evaluation of Climate Models., in *Climate Change 2013:The Physical Science Basis. Contribution of Working Group I to the Fifth Assessment Report of the Intergovernmental Panel on Climate Change*, edited by T. F. Stocker, et al., pp. 741–866, Cambridge University Press, Cambridge, U.K.

Forster, P. M., T. Andrews, P. Good, J. M. Gregory, L. S. Jackson, and M. Zelinka (2013), Evaluating adjusted forcing and model spread for historical and future scenarios in the CMIP5 generation of climate models, *J. Geophys. Res.*, *118*, 1139–1150, doi:10.1002/jgrd.50174.

Gillett, N. P., M. F. Wehner, S. F. B. Tett, and A. J. Weaver (2004), Testing the linearity of the response to combined greenhouse gas and sulfate aerosol forcing, *Geophys. Res. Lett.*, *31*(L13201), doi:10.1029/2004GL020111.

Gillett, N. P., G. M. Flato, J. F. Scinocca, and K. von Salzen (2012), Improved constraints on 21st-century warming derived using 160 years of temperature observations, *Geophys. Res. Lett.*, *39*, L01704, doi:10.1029/2011GL050226.

Gillett, N. P., V. K. Arora, D. Matthews, and M. R. Allen (2013), Constraining the ratio of global warming to cumulative $CO_2$ emissions using CMIP5 simulations, *J. Climate*, *26*, 6844–6858.

Hannart, A., A. Ribes, and P. Naveau (2014), Optimal fingerprinting under multiple sources of uncertainty, *Geophys. Res. Lett.*, *41*(4), 1261–1268, doi:10.1002/2013GL058653.

Harris, G., D. Sexton, B. Booth, M. Collins, J. Murphy, and M. . Webb (2006), Frequency distributions of transient regional climate change from perturbed physics ensembles of general circulation model simulations, *Clim. Dynam.*, *27*, 357–375.

Hasselmann, K. (1997), Multi-pattern fingerprint method for detection and attribution of climate change, *Clim. Dynam.*, *13*, 601–612.

Hegerl, G., and F. Zwiers (2011), Use of models in detection and attribution of climate change, *WIREs Clim. Change*, *2*(4), 570–591.

Hegerl, G. C., et al. (2000), Optimal detection and attribution of climate change: sensitivity of results to climate model differences, *Clim. Dynam.*, *16*(10-11), 737–754.

Hegerl, G. C., et al. (2007), Understanding and attributing climate change, in *Climate Change 2007: The Physical Science Basis. Contribution of Working Group I to the Fourth Assessment Report of the Intergovernmental Panel on Climate Change*, edited by S. Solomon, et al. , pp. 663–745, Cambridge University Press, Cambridge, U.K.

Huntingford, C., P. A. Stott, M. R. Allen, and F. H. Lambert (2006), Incorporating model uncertainty into attribution of observed temperature change, *Geophys. Res. Lett.*, *33*, L05710, doi:10.1029/2005GL0248312.

Jones, G. S., P. A. Stott, and N. Christidis (2013), Attribution of observed historical near surface temperature variations to anthropogenic and natural causes using CMIP5 simulations, *J. Geophys. Res.*, *118*, 4001–4024, doi: 10.1002/jgrd.50239.

Kettleborough, J. A., B. B. B. Booth, P. A. Stott, and M. R. Allen (2007), Estimates of uncertainty in predictions of global mean surface temperature, *J. Climate*, *20*(5), 843–855.

Knutti, R. (2010), The end of model democracy?, *Climatic Change*, *102*, 395–404.

Lamarque, J. F., et al. (2010), Historical (1850-2000) gridded anthropogenic and biomass burning emissions of reactive gases and aerosols: methodology and application, *Atmos. Chem. Phys.*, *10*(15), 7017–7039.

Ledoit, O., and M. Wolf (2004), A well-conditioned estimator for large-dimensional covariance matrices, *J. Multivariate. Anal.*, *88*, 365–411.

Lee, T. C. K., F. W. Zwiers, G. C. Hegerl, X. B. Zhang, and M. Tsao (2005), A Bayesian climate change detection and attribution assessment, *J. Climate*, *18*(13), 2429–2440.

Mardia, K. V., J. T. Kent, and J. M. Bibby (1979), *Multivariate Analysis*, Academic Press, London, U.K.

Meehl, G. A., R. Moss, K. E. Taylor, V. Eyring, R. J. Stouffer, S. Bony, and B. Stevens (2014), Climate Model Intercomparison: Preparing for the Next Phase, *EOS*, *95*(9), 77–78, doi:10.1002/2014EO090001.

Mitchell, J. F. B., D. J. Karoly, G. C. Hegerl, F. W. Zwiers, M. R. Allen, and J. Marengo (2001), Detection of climate change and attribution of causes, in *Climate change 2001: The scientific basis. Contribution of Working Group I to the Third Assessment Report of the Intergovernmental Panel on Climate Change*, edited by J. T. Houghton, et al., pp. 695–738, Cambridge University Press, Cambridge, U.K.

Morice, C. P., J. J. Kennedy, N. A. Rayner, and P. D. Jones (2012), Quantifying uncertainties in global and regional temperature change using an ensemble of observational estimates: the HadCRUT4 dataset, *J. Geophys. Res.*, *117*, D08101, doi:10.1029/2011JD017187.

Myhre, G., et al. (2013), Anthropogenic and Natural Radiative Forcing., in *Climate Change 2013:The Physical Science Basis. Contribution of Working Group I to the Fifth Assessment Report of the Intergovernmental Panel on Climate Change*, edited by T. F. Stocker, et al., pp. 659–740, Cambridge University Press, Cambridge, U.K.

Ribes, A., S. Planton, and L. Terray (2013), Application of regularised optimal fingerprinting to attribution. Part I: method, properties and idealised analysis, *Clim. Dynam.*, *41*(11-12), 2817–2836.

Ribes, A., and L. Terray (2013), Application of regularised optimal fingerprinting to attribution. Part II: application to global near-surface temperature, *Clim. Dynam.*, *41*(11-12), 2837–2853.

Ribes, A., N. P. Gillett, and F. W. Zwiers (2015), Designing detection and attribution simulations for CMIP6 to optimize the estimation of greenhouse-gas induced warming, *J. Clim.*, *28*, 3435–3438.

Santer, B. D., et al. (2012), Incorporating model quality information in climate change detection and attribution studies, *P. Natl. Acad. Sci. USA*, *106*(35), 14,778–14,783.

Shindell, D. T., et al. (2013), Radiative forcing in the ACCMIP historical and future climate simulations, *Atmos. Chem. Phys.*, *13*(6), 2939–2974.

Stott, P., P. Good, G. S. Jones, N. Gillett, and E. Hawkins (2013), The upper end of climate model temperature projections is inconsistent with past warming, *Environ. Res. Lett.*, *8*, 014024, doi:10.1088/1748-9326/8/1/014024.

Stott, P. A., and C. E. Forest (2007), Ensemble climate predictions using climate models and observational constraints, *Philos. T. R. Soc. S-A*, *365*, 2029–2052.

Stott, P. A., and G. S. Jones (2012), Observed 21st century temperatures further constrain decadal predictions of future warming, *Atmos. Sci. Lett.*, *13*(3), 151–156.

Stott, P. A., and J. A. Kettleborough (2002), Origins and estimates of uncertainty in predictions of twenty-first century temperature rise, *Nature*, *416*, 723–726.

Stott, P. A., and S. F. B. Tett (1998), Scale-dependent detection of climate change, *J. Climate*, *11*(12), 3282–3294.

Stott, P. A., M. R. Allen, and G. S. Jones (2003), Estimating signal amplitudes in optimal fingerprinting, part  II: application to general circulation models, *Clim. Dynam.*, *21*(5-6), 493–500.

Stott, P. A., et al. (2006), Observational constraints on past attributable warming and predictions of future  global warming, *J. Climate*, *19*(13), 3055–3069.

Taylor, K. E., R. J. Stouffer, and G. A. Meehl (2012), An overview of CMIP5 and the experiment design, *B. Am. Meteorol. Soc.*, *93*(4), 485–498.

Tett, S. F. B., P. A. Stott, M. R. Allen, W. J. Ingram, and J. F. B. Mitchell (1999), Causes of twentieth-century temperature change near the Earth's surface, *Nature*, *399*, 569–572.

Tett, S. F. B., et al. (2002), Estimation of natural and anthropogenic contributions to 20th Century temperature change, *J. Geophys. Res.*, *107*(D16), 4306, doi:10.1029/2000JD000028.

Wilcox, L. J., E. J. Highwood, B. B. B. Booth, and K. S. Carslaw (2015), Quantifying sources of inter-model diversity in the cloud albedo effect, *Geophys. Res. Lett.*, *42*, 1568–1575, doi:10.1002/2015GL063301.

Wilcox, L. J., E. J. Highwood, and N. J. Dunstone (2013), The influence of anthropogenic aerosol on multi-decadal variations of historical global climate, *Environ. Res. Lett.*, *8*(2), 024033, doi:10.1088/1748-9326/8/2/024033.

Wu, P., N. Christidis, and P. Stott (2013), Anthropogenic impact on Earth's hydrological cycle, *Nat. Clim. Chang.*, *3*, 807–810.

Table 1: Experiment definitions used in this study. piControl, historical, historicalGHG and historicalNat follow CMIP5 nomenclature (*Taylor et al.*, 2012). * - Well mixed greenhouse gases include carbon dioxide, methane, nitrous oxides and CFCs/HCFCs. Some models also include variations in ozone concentration in the experiment (Table 2). †- anthropogenic forcings include well mixed greenhouse gases, ozone, sulphate aerosols, carbonaceous aerosols and land use changes. Experiments historicalOA, historicalGOA and historicalOAN are not CMIP5 experiments. They are defined for this study as simple linear combinations of the historical, historicalGHG and historicalNat experiments.

| Experiment | Definition |
| --- | --- |
| piControl | Constant preindustrial forcing factors |
| historicalGHG | Variations in well-mixed greenhouse gas concentrations* |
| historicalNat | Variations in solar irradiance and volcanic stratospheric aerosols |
| historical | Variations in historic anthropogenic† and natural radiative forcings |
| historicalOA | Variations in other (non well-mixed greenhouse gas) anthropogenic factors. |
| historicalGOA | Variations in all anthropogenic factors. |
| historicalOAN | Variations in other anthropogenic and natural factors. |

Table 2: Details of CMIP5 models used in analysis. The 15 models (A-O) with historical, historicalNat and historicalGHG experiments are listed together with their transient climate response (TCR) (*Forster et al.*, 2013); whether, yes (Y) or no (N), indirect aerosol effects are included (SI); whether ozone influences are included in the historicalGHG experiment (O3); whether anthropogenic land use changes are included in the historical experiment (LU); and the number of initial condition ensembles for each experiment. In addition to models A-O, a further 8 models were used that had over 500 years of piControl avalable; ACCESS1-3, CESM1-BGC, GFDL-ESM2G, MIROC5, MPI-ESM-LR, MPI-ESM-MR, MPI-ESM-P and inmcm4 - a total of 23 models. †- Model K simulated the radiative effects of stratospheric volcanic aerosol by varying incoming total solar irradiance in the historical and historicalNat experiments. Model K also included land use changes in the historicalGHG experiment.

| Model index | Model | TCR (K) | SI | O3 | LU | Ensemble members historical/ historicalNat/ historicalGHG |
|---|---|---|---|---|---|---|
| A | BNU-ESM | 2.6 | N | N | N | 1/1/1 |
| B | CCSM4 | 1.8 | N | N | Y | 6/4/3 |
| C | CNRM-CM5 | 2.1 | Y | Y | N | 10/6/6 |
| D | CSIRO-Mk3-6-0 | 1.8 | Y | N | N | 10/5/5 |
| E | CanESM2 | 2.4 | Y | N | Y | 5/5/5 |
| F | GFDL-CM3 | 2.0 | Y | Y | Y | 5/3/3 |
| G | GFDL-ESM2M | 1.3 | N | N | Y | 1/1/1 |
| H | GISS-E2-H | 1.7 | Y | N | Y | 6/5/5 |
| I | GISS-E2-R | 1.5 | Y | N | Y | 6/5/5 |
| J | HadGEM2-ES | 2.5 | Y | N | Y | 4/4/4 |
| K | IPSL-CM5A-LR† | 2.0 | Y | Y | Y | 6/3/3 |
| L | MIROC-ESM | 2.2 | Y | Y | Y | 3/3/3 |
| M | MRI-CGCM3 | 1.6 | Y | Y | Y | 3/1/1 |
| N | NorESM1-M | 1.4 | Y | Y | Y | 3/1/1 |
| O | bcc-csm1 | 1.7 | N | N | N | 3/1/1 |

Table 3: Cross correlations between the different patterns used in the optimal detection analyses and the observations, HadCRUT4. Shown are Pearson correlation coefficients between pattern *a* and pattern *b*, as used in the regression (Equation 1) following filtering and after the data has been 'whitened'.

| Model | HadCRUT4 | | | GOAN | | G | *a* |
|---|---|---|---|---|---|---|---|
| | GOAN | G | N | G | N | N | *b* |
| A | 0.82 | 0.83 | 0.12 | 0.87 | 0.15 | 0.05 | |
| B | 0.89 | 0.87 | 0.16 | 0.95 | 0.10 | -0.08 | |
| C | 0.87 | 0.85 | 0.11 | 0.89 | 0.10 | -0.10 | |
| D | 0.76 | 0.82 | 0.14 | 0.56 | 0.56 | -0.08 | |
| E | 0.82 | 0.85 | -0.18 | 0.87 | -0.10 | -0.43 | |
| F | 0.65 | 0.82 | 0.09 | 0.52 | 0.42 | -0.09 | |
| G | 0.68 | 0.75 | -0.03 | 0.69 | 0.13 | -0.14 | |
| H | 0.83 | 0.86 | 0.18 | 0.86 | 0.26 | -0.08 | |
| I | 0.83 | 0.84 | 0.07 | 0.83 | 0.13 | -0.16 | |
| J | 0.58 | 0.83 | 0.00 | 0.51 | 0.23 | -0.16 | |
| K | 0.86 | 0.84 | 0.02 | 0.93 | 0.06 | -0.16 | |
| L | 0.85 | 0.84 | 0.37 | 0.79 | 0.37 | 0.16 | |
| M | 0.64 | 0.78 | -0.09 | 0.53 | 0.24 | -0.17 | |
| N | 0.89 | 0.79 | 0.14 | 0.83 | 0.11 | -0.08 | |
| O | 0.87 | 0.77 | 0.11 | 0.89 | 0.12 | -0.03 | |
| Mod.Avg. | 0.87 | 0.85 | 0.11 | 0.88 | 0.20 | -0.16 | |

Table 4: Signal to noise ratios (SNRs) for ensemble mean patterns used in the optimal regression, after filtering and projected onto first 40 EOFs. The historical all forcings (GOAN), well mixed greenhouse gases (G) and natural (N) forcing responses are directly included in the regressions. The OA, GOA and OAN signals are not directly involved in the regressions, but calculated here to demonstrate the relative strength of the inferred patterns.

| | GOAN | G | N | OA | GOA | OAN |
|---|---|---|---|---|---|---|
| A | 3.04 | 3.62 | 1.34 | 1.14 | 2.20 | 1.21 |
| B | 6.40 | 4.71 | 1.43 | 0.75 | 4.11 | 1.12 |
| C | 5.41 | 6.53 | 1.39 | 1.94 | 3.40 | 2.67 |
| D | 4.28 | 6.85 | 1.63 | 3.34 | 2.02 | 4.71 |
| E | 5.53 | 8.27 | 1.88 | 1.92 | 4.31 | 3.10 |
| F | 3.95 | 7.57 | 1.77 | 3.84 | 2.24 | 5.23 |
| G | 2.29 | 2.49 | 1.68 | 1.18 | 1.92 | 1.34 |
| H | 4.66 | 6.70 | 1.36 | 1.93 | 3.07 | 2.72 |
| I | 4.33 | 6.06 | 1.52 | 1.85 | 3.02 | 2.57 |
| J | 2.97 | 6.68 | 1.93 | 3.11 | 2.27 | 4.06 |
| K | 7.32 | 7.63 | 1.55 | 1.87 | 4.36 | 2.76 |
| L | 3.28 | 5.03 | 1.51 | 1.93 | 2.17 | 2.27 |
| M | 2.22 | 2.69 | 0.89 | 1.42 | 1.17 | 1.99 |
| N | 2.91 | 2.89 | 1.01 | 1.16 | 1.61 | 1.52 |
| O | 4.34 | 3.04 | 1.17 | 0.96 | 2.29 | 1.17 |
| Mod.Avg. | 12.75 | 16.86 | 3.45 | 4.72 | 7.97 | 7.13 |

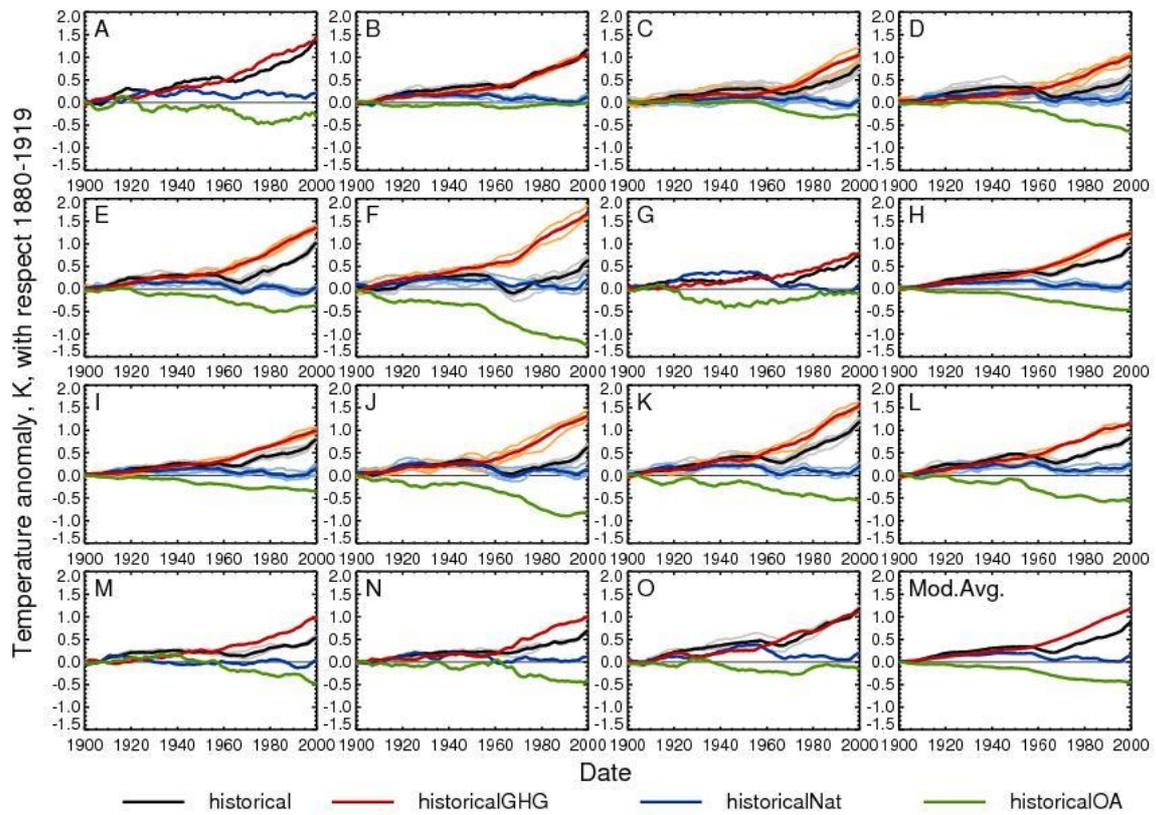

Figure 1: Global, ten year running mean, near surface temperature variations for each model, anomalies relative to the 1880-1919 period. Model ensemble means shown as thick dark lines and individual ensemble members as thin light lines. The 'Mod.Avg.' is the multi-model average, calculated as the mean of the model ensemble means. The historical, historicalGHG and historicalNat experiments are the CMIP5 experiments, with historicalOA inferred from the difference between historical and the sum of historicalGHG and historicalNat.

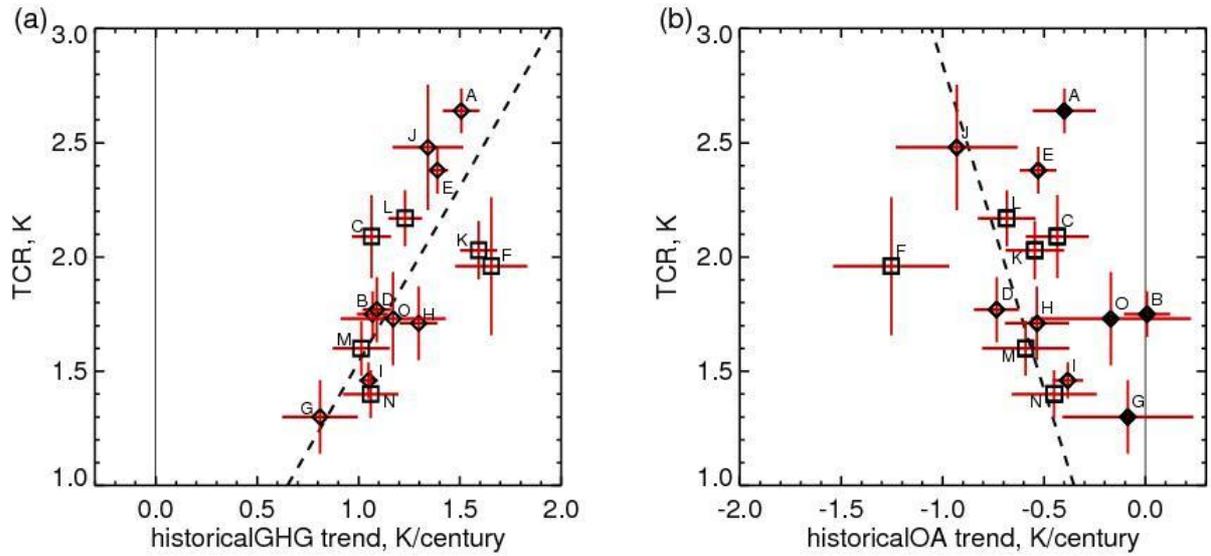

Figure 2: Comparison of Transient Climate Response (TCR), Table 2, with the mean historicalGHG (a) and historicalOA (b) temperature trends over the 1906-2005 period. The 2.5-97.5% uncertainty ranges in TCR are estimated from the variability in 20 year means separated by 70 years in each model's piControl. The 2.5-97.5% ranges in temperature trends are estimated from the variability of 100 year trends in the model's piControl, scaled to account for the number of ensemble members. Diamond symbols signify models which do not include ozone forcing as part of the historicalGHG experimental design, and square symbols where ozone is included as a forcing in historicalGHG. Infilled symbols, in (b), represent models which do not include aerosol indirect effects in the historical simulations. The dashed lines in both panels are linear regression lines through the data, passing through 0 in both axes, and are included as a guide.

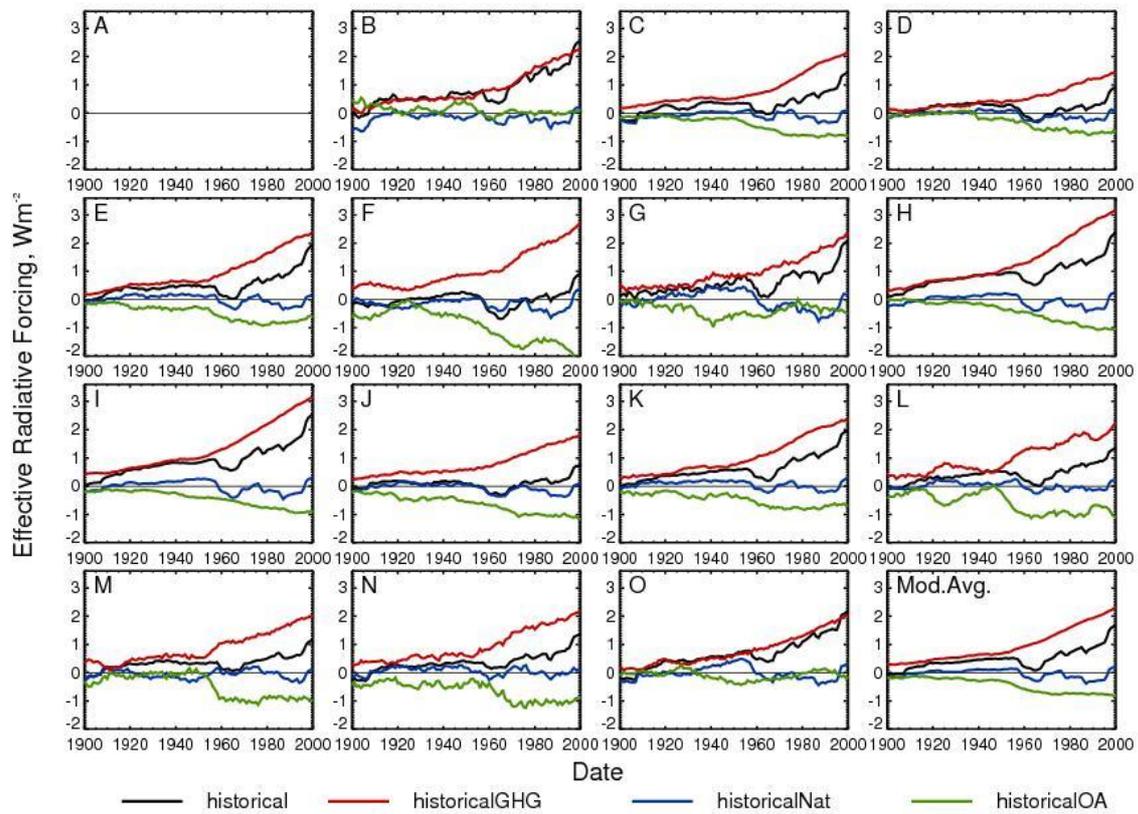

Figure 3: Global, ten year running mean effective radiative forcing (ERF), Wm$^{-2}$, as calculated by *Forster et al.* (2013) - described as 'adjusted forcing, AF' in that study. ERF is calculated relative to the model's own piControl. The 'Mod.Avg.' is the multi-model average, calculated as the mean across the models. For model A the required data to estimate ERF was not available.

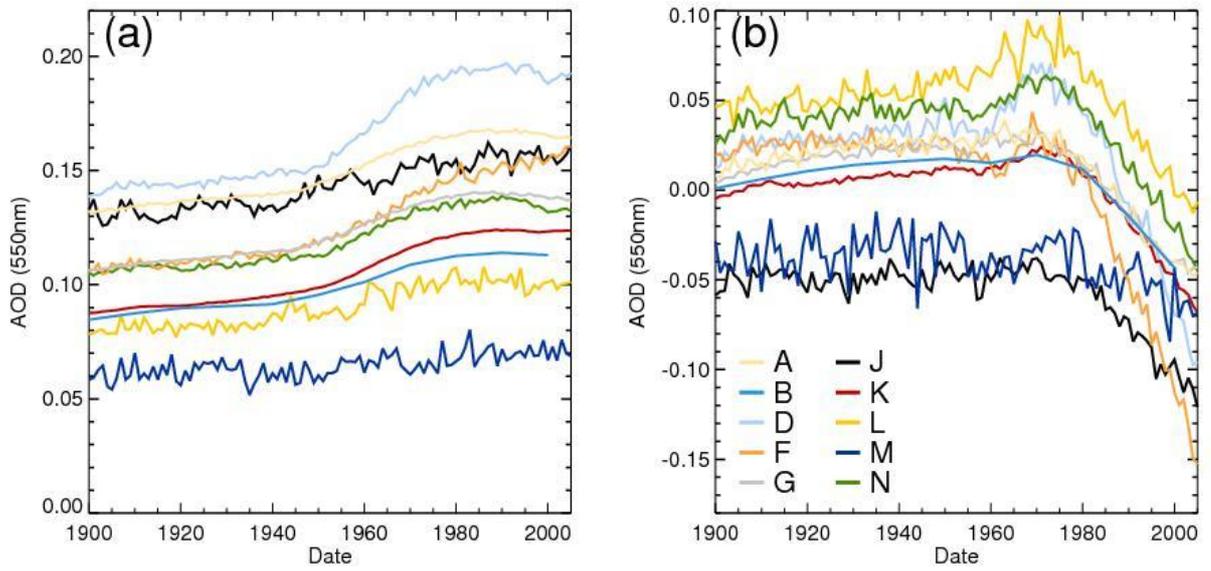

Figure 4: Aerosol optical depth (AOD) at 550nm for the models with available diagnostics. (a) global mean, (b) difference between region covering North America to Europe (120ºW-50ºE and 30ºN-70ºN) and the South East Asia region (50ºE-150ºE and 0ºN-30ºN).

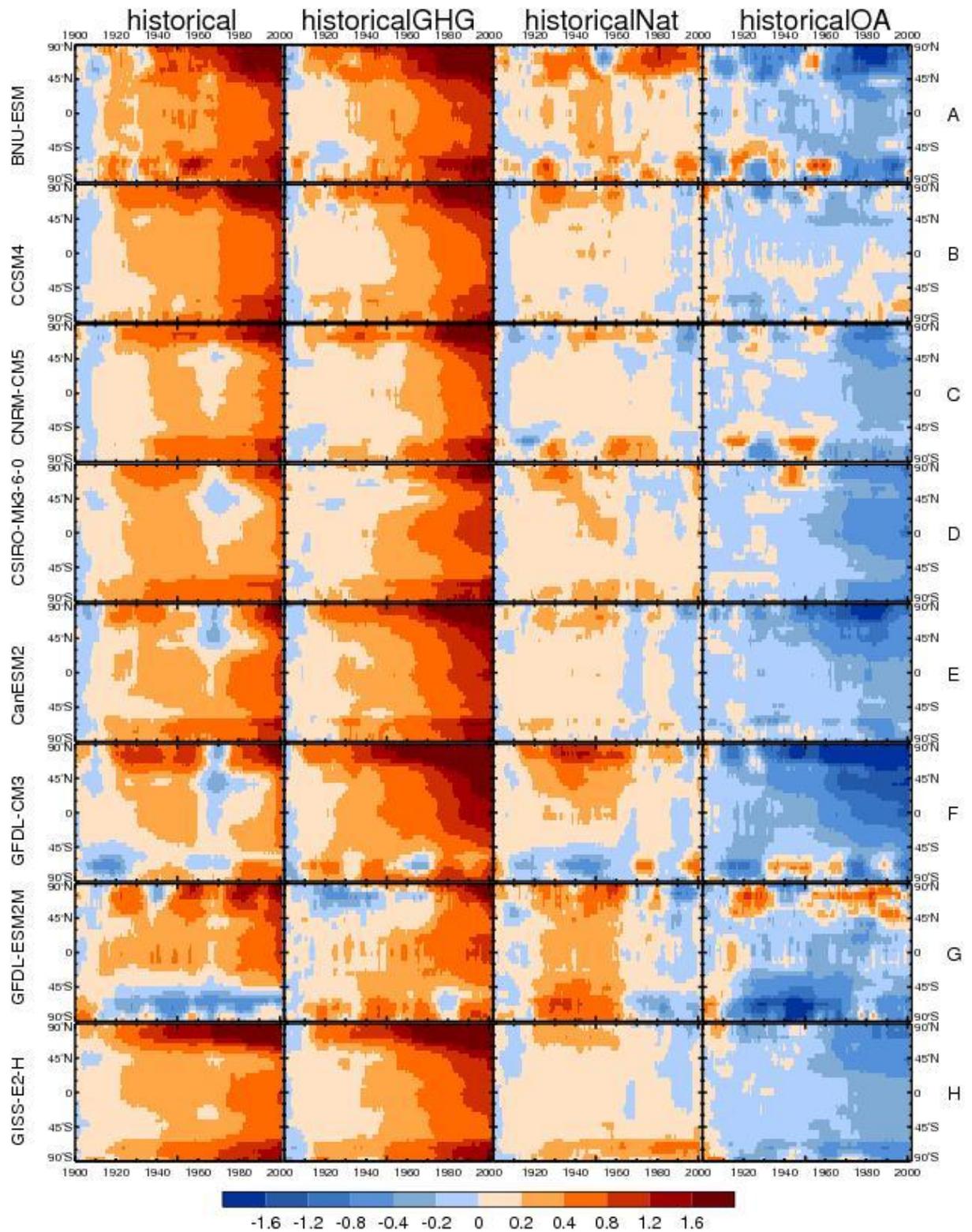

Figure 5: Zonal, 10 year running means, near surface temperatures for the ensemble means of historical, historical-GHG and historicalNat (first three columns) and historicalOA (final column). Models A to H shown. Temperatures (K) given as anomalies relative to 1880-1919 period.

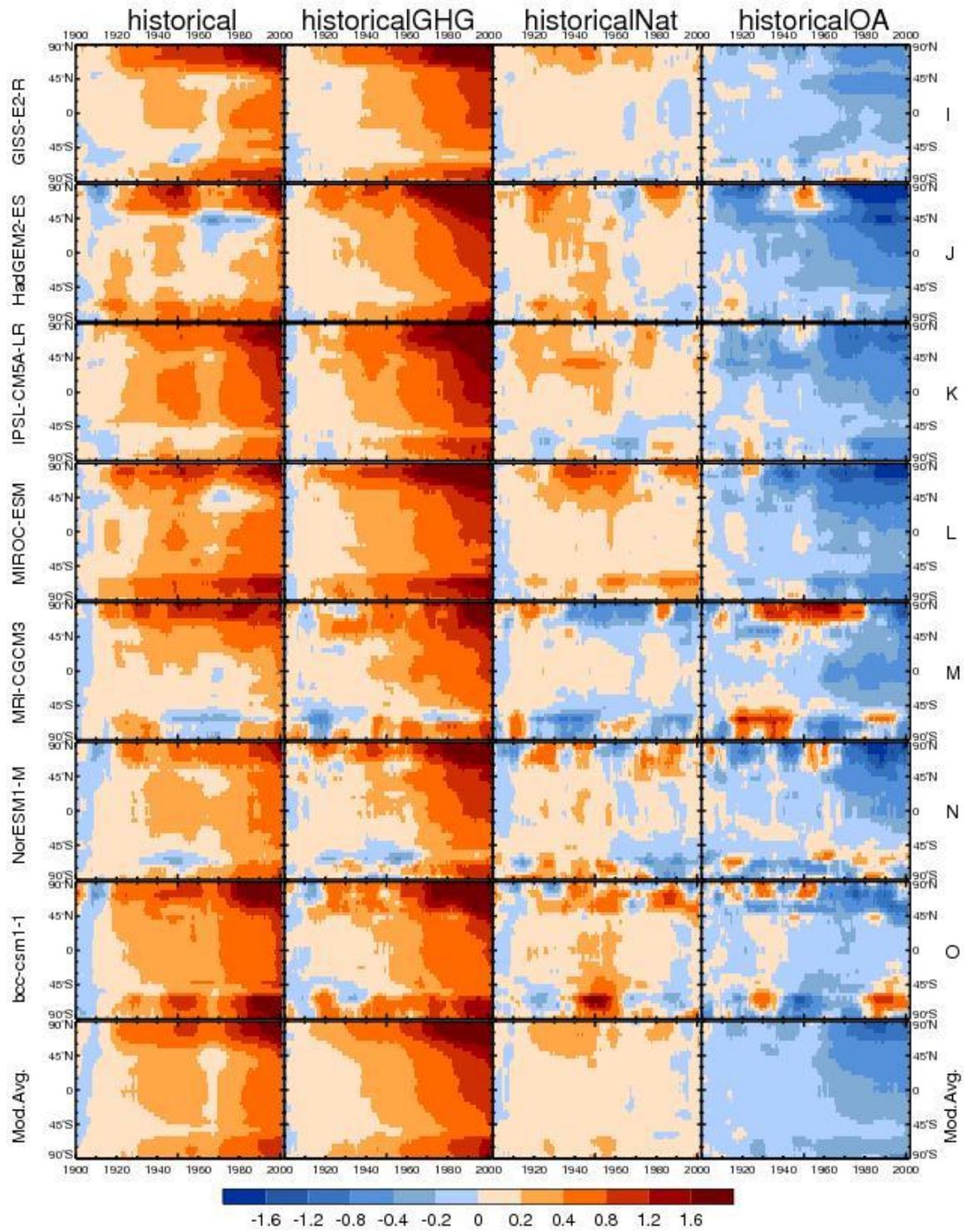

Figure 6: As Figure 5 but for models I to O and 'Mod.Avg.'.

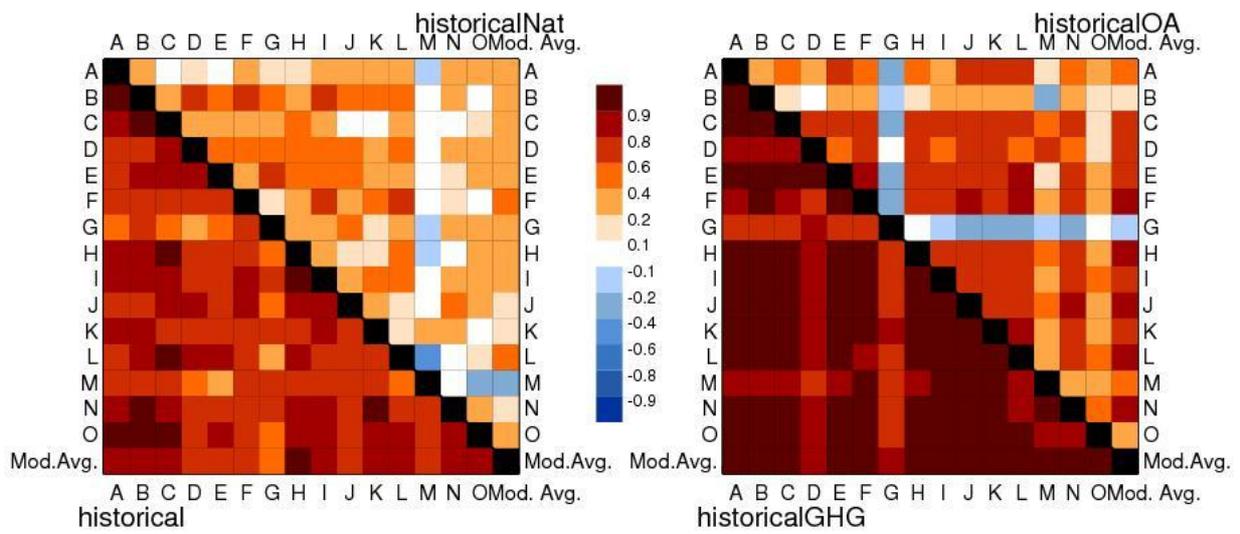

Figure 7: Area weighted spatio-temporal cross correlations between models, for the four experiments historical (bottom left of left panel), historicalNat (top right of left panel), historicalGHG (bottom left of right panel) and historicalOA (top right of right panel). The data used in the correlations are sampled from Figures 5 and 6 for the 1906-2005 period using independent 10 year means.

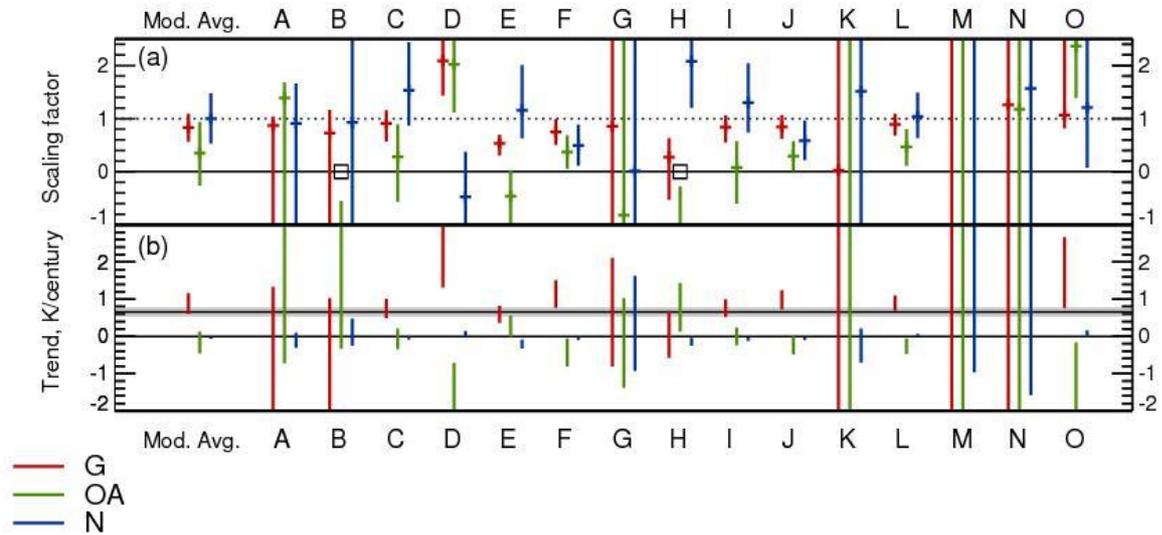

Figure 8: Optimal detection analysis for G, OA and N. Scaling factors (a) and scaled linear trends (b) for the standard analysis for the 1906-2005 period, using 10 year means, spatial meaning of spherical harmonics (T4) and an EOF truncation of 40. Uncertainties given as $5-95\%$ ranges. Squares in panel (a) indicate where the residual of regression fails an F-Test when compared with measure of internal variability (*Allen and Stott*, 2003). The horizontal line and shaded band in panel (b) are the observed trend of 0.65K/century, together with estimate of $5-95\%$ uncertainty range due to internal variability, estimated from the CMIP5 piControl variability (*Tett et al.*, 2002).

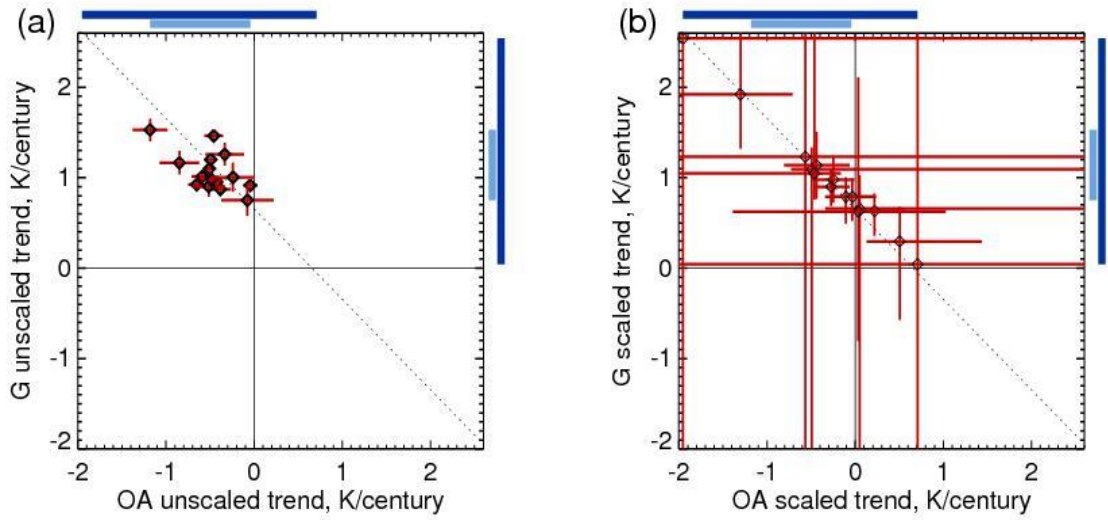

Figure 9: Relationship between G and OA trends for 1906-2005 period (K/decade) for the G, OA, N analysis. (a): relationship between unscaled OA and unscaled G trends with uncertainties representing 5-95% range due to internal variability. (b) relationship between scaled OA and scaled G trends, with uncertainties representing 5-95% range due to scaling factor ranges. Bars on axis represent min-max range of best estimates before scaling - inner lighter coloured bar - and after scaling - outer darker coloured bar. Dotted line used as a guide to show where sum of trends equals the observed trend of 0.65K/century.

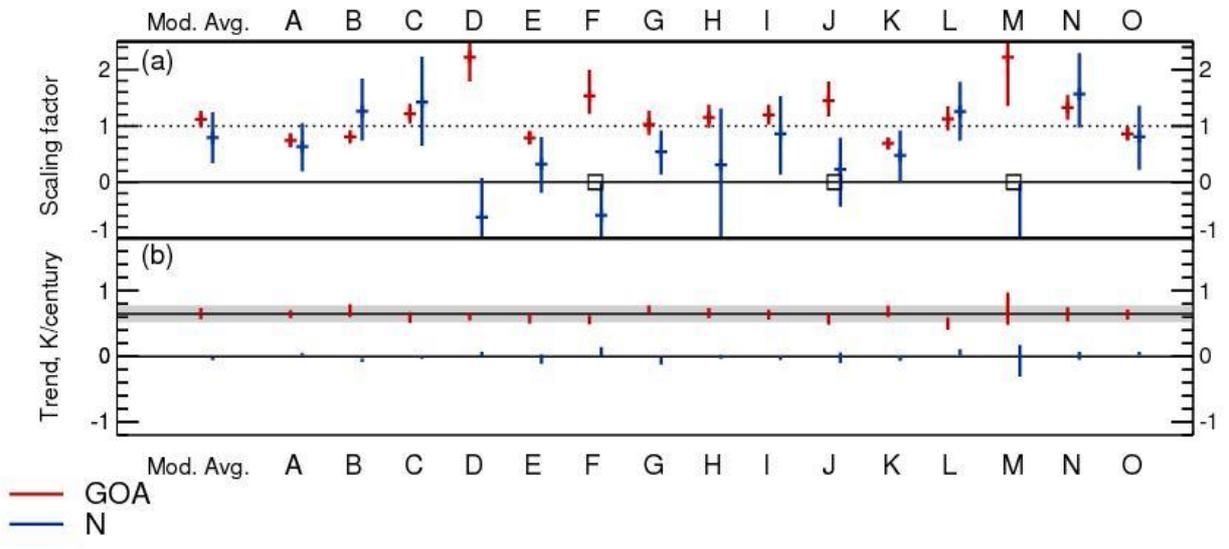

Figure 10: As Figure 8 but for two way regression of GOA and N. Note the smaller trend range displayed in panel (b) than in Figure 8(b)

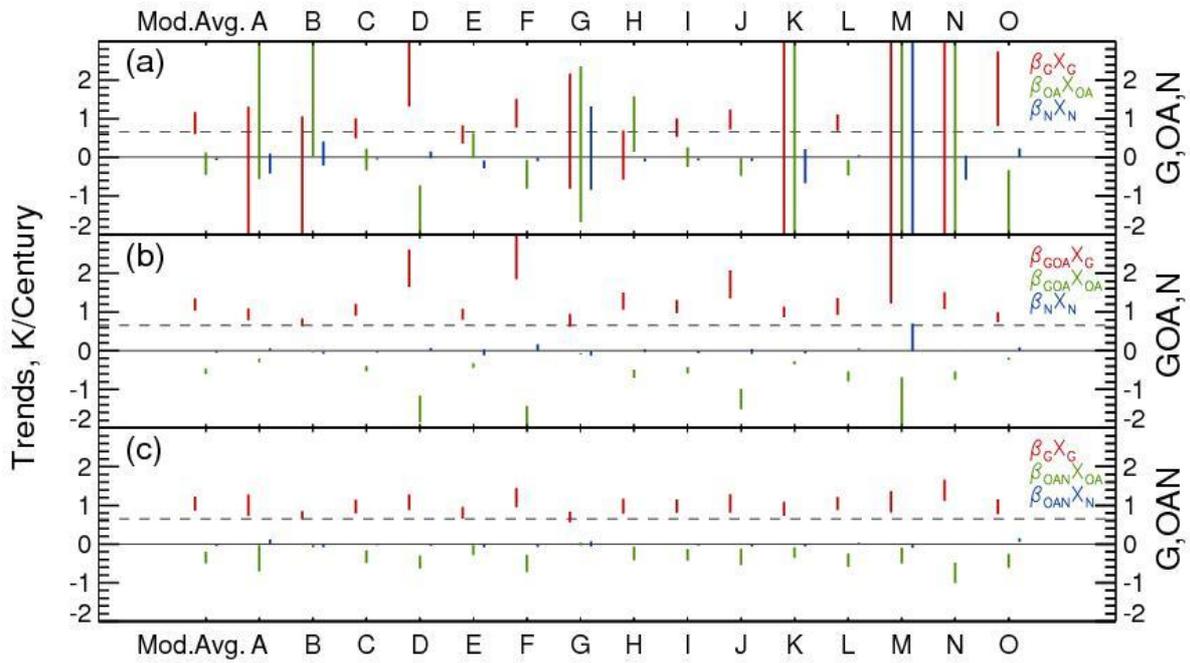

Figure 11: Comparisons of estimated contributions of G, OA and N (red, green and blue bars respectively) to the scaled trends deduced from the different optimal detection analyses. (a) G,OA,N analysis: Trends of $\beta_G X_G$, $\beta_{OA} X_{OA}$ and $\beta_N X_N$; (b) GOA,N analysis: Trends of $\beta_{GOA} X_G$, $\beta_{GOA} X_{OA}$ and $\beta_N X_N$; (c) G,OAN analysis: Trends of $\beta_G X_G$, $\beta_{OAN} X_{OA}$ and $\beta_{OAN} X_N$. Dashed line is the 1906-2005 HadCRUT4 trend.

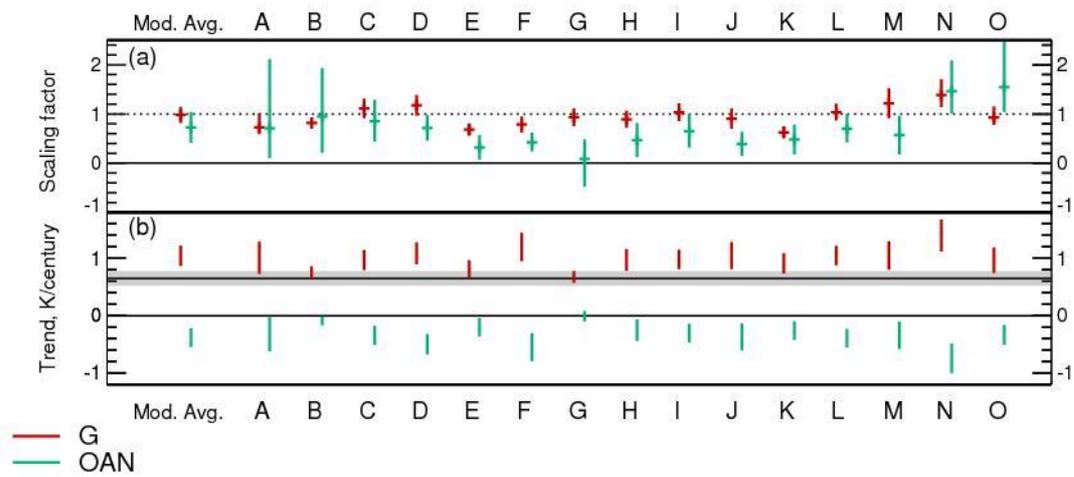

Figure 12: As Figure 8 but for the two signal regression of G, OAN.

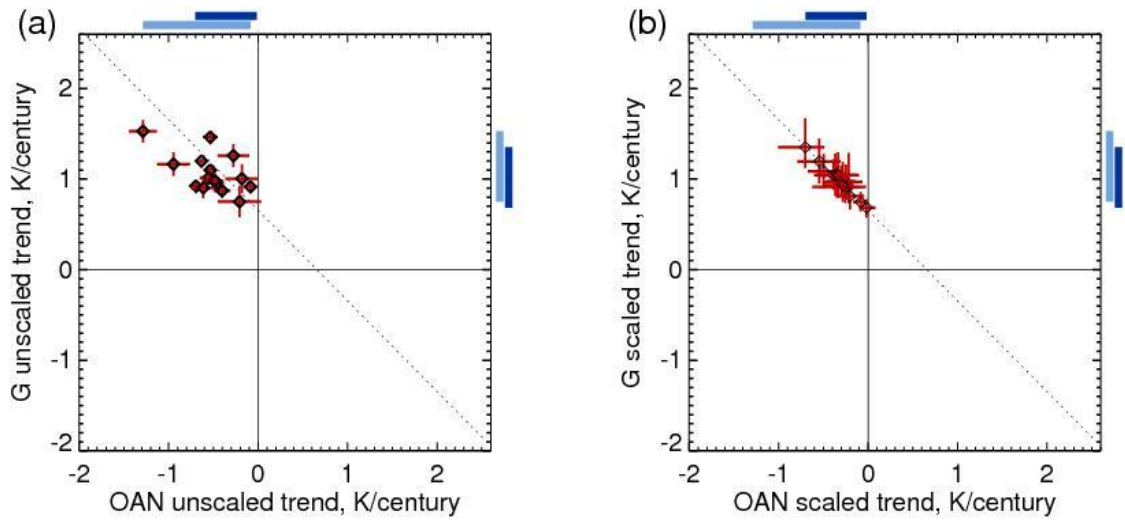

Figure 13: Relationship between G and OAN trends for 1906-2005 period (K/decade) for G, OAN analysis. (a): relationship between unscaled OAN trend and unscaled G trend (b) relationship between scaled OAN trend and scaled G trend. See Figures 12and 9.

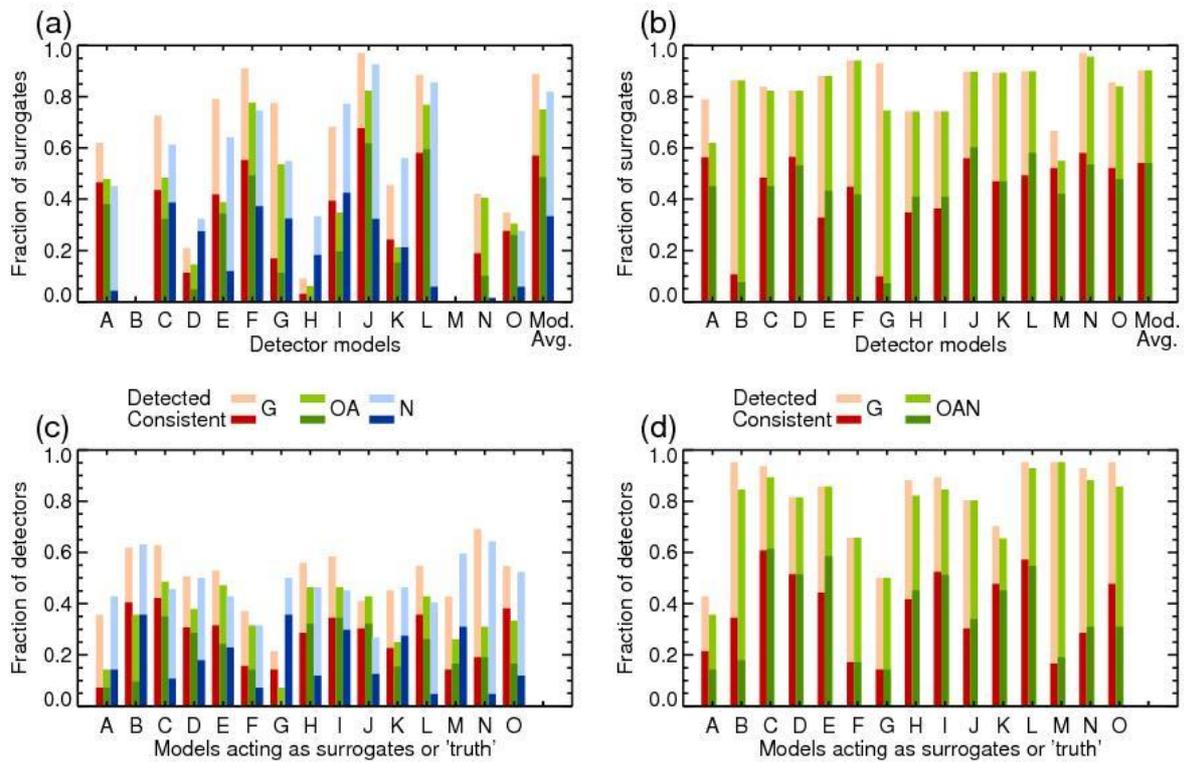

Figure 14: Summary of perfect model study results for the G,OA,N analysis (a,c) and the G,OAN analysis (b,d). Shown are fractions of cases where the signal is detected (light colour bars) and where the scaled signal trend is consistent with the expected trend of the signal for the model being used a surrogate observation (dark colour bars) - both where the residual passes the consistency test. Panels (a,b) show, for each predictor (or detector) model, the fractions of surrogate observations - up to 72, drawn from historical simulations from the other 14 models. Panels (c,d) show for each model that is used as a predictand or surrogate observation (our 'truth' model if you will) the fraction of predictor models - up to 14 times the number of ensemble members for that predictand model.

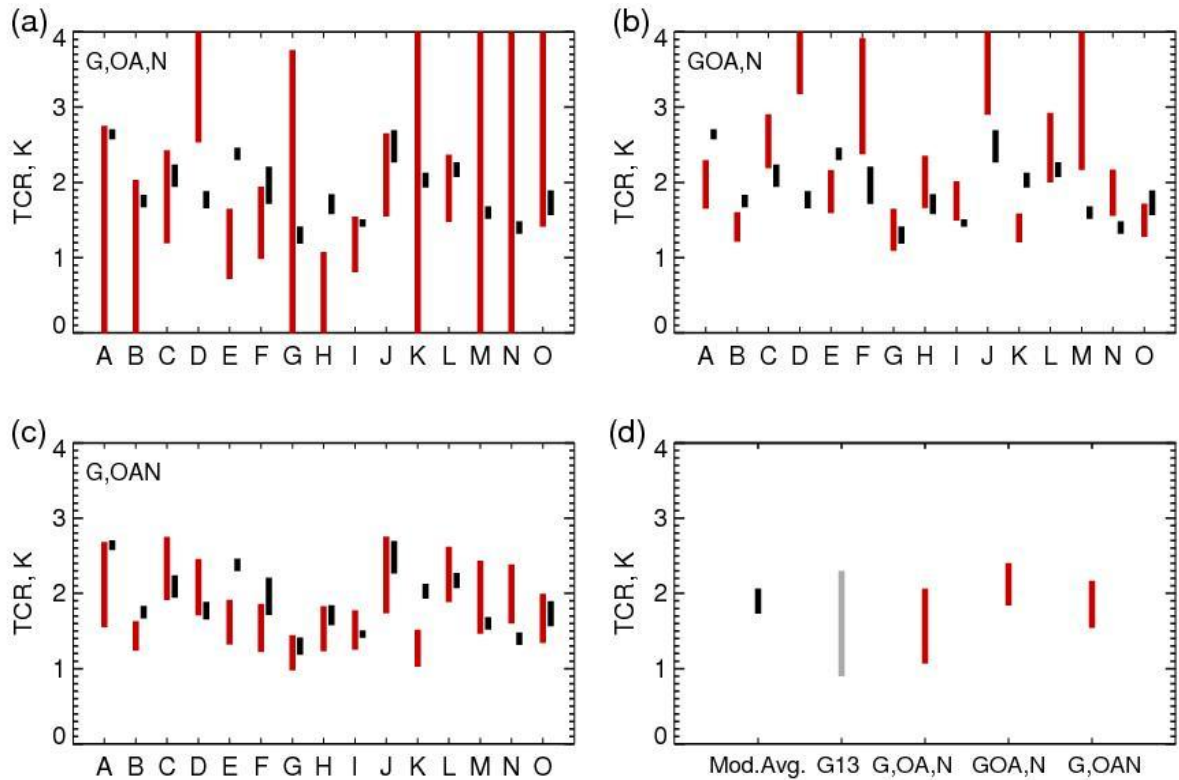

Figure 15: Transient climate response (TCR) deduced from each of the optimal detection analyses; (a) G,OA,N, (b) GOA,N and (c) G,OAN. Uncertainty ranges, $5-95\%$, for TCR for each model before scaling (black lines) calculated from variability from each model's piControl. Uncertainty ranges, $5-95\%$, after scaling (red lines) calculated from regression scaling factors. (d) The multi-model average (Mod.Avg.) of TCR before and after scaling for different analyses, as well as the scaled multi-model average TCR from *Gillett et al.* (2013) (G13). The uncertainty range of the Mod.Avg. unscaled TCR estimated by boot strapping the means from each of the models, to give an estimate of the 'true' multi-model range.

# Supporting Information for "Uncertainties in the attribution of greenhouse gas warming and implications for climate prediction"

Gareth S. Jones*†, Peter A. Stott†and John F. B. Mitchell†

†- Met Office Hadley Centre, Exeter, UK

**Contents of this file**

**Introduction** Included in this supporting information is text summarising the linear transformations used in the different optimal detection analyses described in the study and additional figures showing sensitivity of results to alternative choices.

---

*Corresponding author:Gareth S Jones, Met Office Hadley Centre, FitzRoy Road, Exeter, EX1 3PB, UK (gareth.s.jones@metoffice.gov.uk)

## Text S1. Linear Transformations

The following summarises how the patterns ($x_i$) and scaling factors ($\beta_i$) for the experiments in the different optimal detection analyses were linearly transformed. See Appendix B in *Tett et al.* (2002) for more details.

There are no transformations for the single pattern regressions of historical, historicalGHG and historicalNat to GOAN, G and N respectively. Similarly no transformation is required for the two way regression of historicalGHG and historicalNat to G and N.

### S1.1. historical, historicalGHG and historicalNat to G, OA and N

$$\left.\begin{aligned}
x_G &= x_{historicalGHG} \\
x_{OA} &= x_{historicalOA} = x_{historical} - x_{historicalGHG} - x_{historicalNat} \\
x_N &= x_{historicalNat}
\end{aligned}\right\} \quad (1)$$

$$\left.\begin{aligned}
\beta_G &= \beta_{historical} + \beta_{historicalGHG} \\
\beta_{OA} &= \beta_{historical} \\
\beta_N &= \beta_{historical} + \beta_{historicalNat}
\end{aligned}\right\} \quad (2)$$

### S1.2. historical and historicalNat to GOA and N

$$\left.\begin{aligned}
x_{GOA} &= x_{historicalGOA} = x_{historical} - x_{historicalNat} \\
x_N &= x_{historicalNat}
\end{aligned}\right\} \quad (3)$$

$$\left.\begin{aligned}
\beta_{GOA} &= \beta_{historical} \\
\beta_N &= \beta_{historical} + \beta_{historicalNat}
\end{aligned}\right\} \quad (4)$$

### S1.3. historical and historicalGHG to G and OAN

$$\left.\begin{aligned}
x_G &= x_{historicalGHG} \\
x_{OAN} &= x_{historicalOAN} = x_{historical} - x_{historicalGHG}
\end{aligned}\right\} \quad (5)$$

$$\left.\begin{aligned}
\beta_G &= \beta_{historicalGHG} + \beta_{historical} \\
\beta_{OAN} &= \beta_{historical}
\end{aligned}\right\} \quad (6)$$

# References

Jones, G. S., P. A. Stott, and N. Christidis (2013), Attribution of observed historical near surface temperature variations to anthropogenic and natural causes using CMIP5 simulations, *Journal of Geophysical Research*, *118*, 4001–4024.

Tett, S. F. B., et. al. (2002), Estimation of natural and anthropogenic contributions to 20th Century temperature change, *Journal of Geophysical Research*, *107*(D16), 4306, doi:10.1029/2000JD000028.

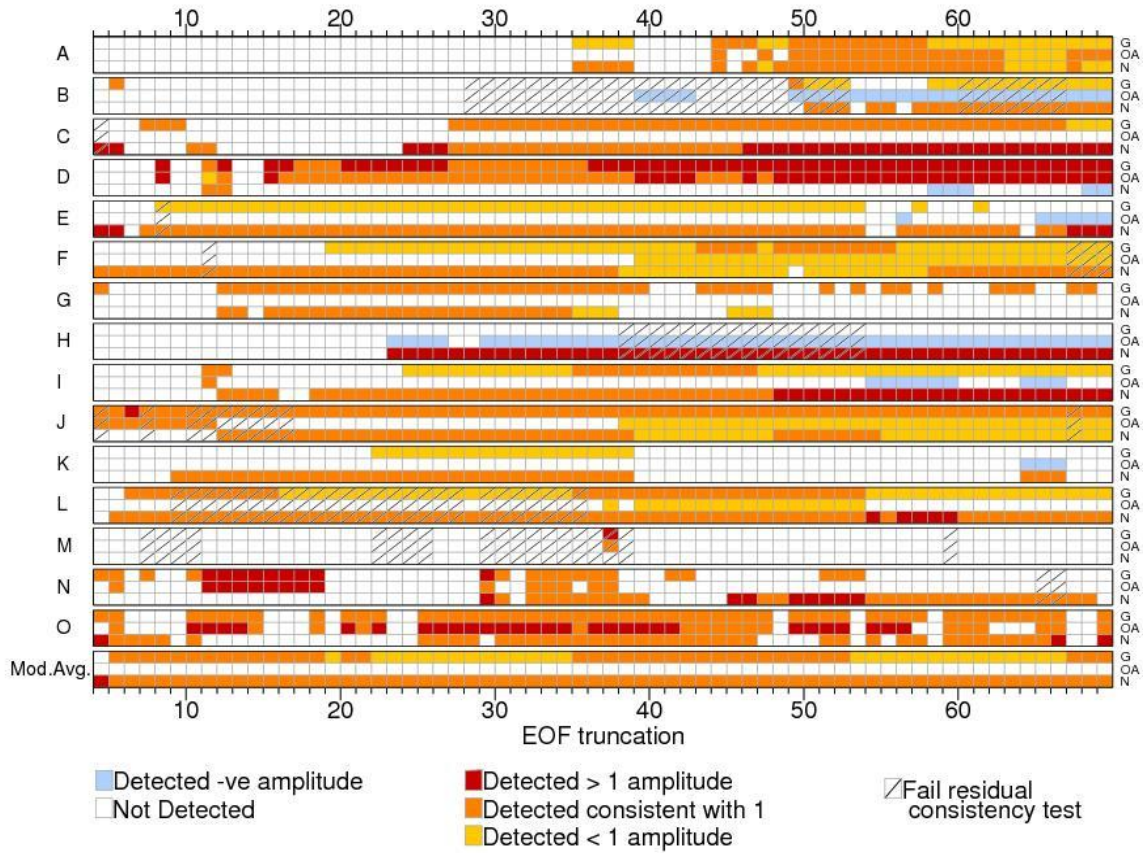

Figure S1: Summary of optimal detection results for G,OA,N regression analysis showing sensitivity to choice of EOF truncation. Analysis on 1906-2005 period for 10year means projected onto T4 spherical harmonics, number of degrees of freedom is 69 [See main text for details]. Instead of showing the scaling factors and their uncertainties for each truncation and each model, as is sometimes done (e.g., *Jones et al.*, 2013) here we show where the scaling factors are detected - coloured regions - and where the residual statistical test fails - cross hatching. The different colours represent where the signals are detected with scaling factors greater than 1, consistent with 1 and less than 1, as well as when the scaling factors are negative.

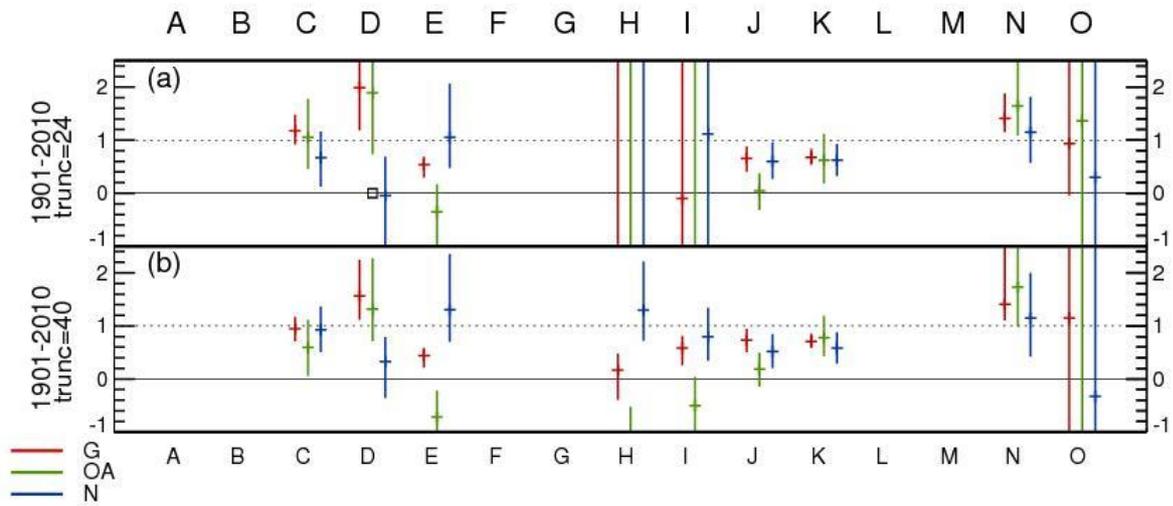

Figure S2: Results for three way regression, G,OA,N, for the 1901-2010 period, for 10 year means projected onto T4 spherical harmonics. The number of models that can cover this range is reduced to nine (C, D, E, H, I, J, K, M and O), with historicalGHG and historicalNat experiments covering this period and historical extended with either historicalExt or rcp45 experiments - as described in *Jones et al.* (2013). (a) scaling factors for EOF truncation of 24, as used in *Jones et al.* (2013) and (b) scaling factors for EOF truncation of 40, as used in the main text of this study.

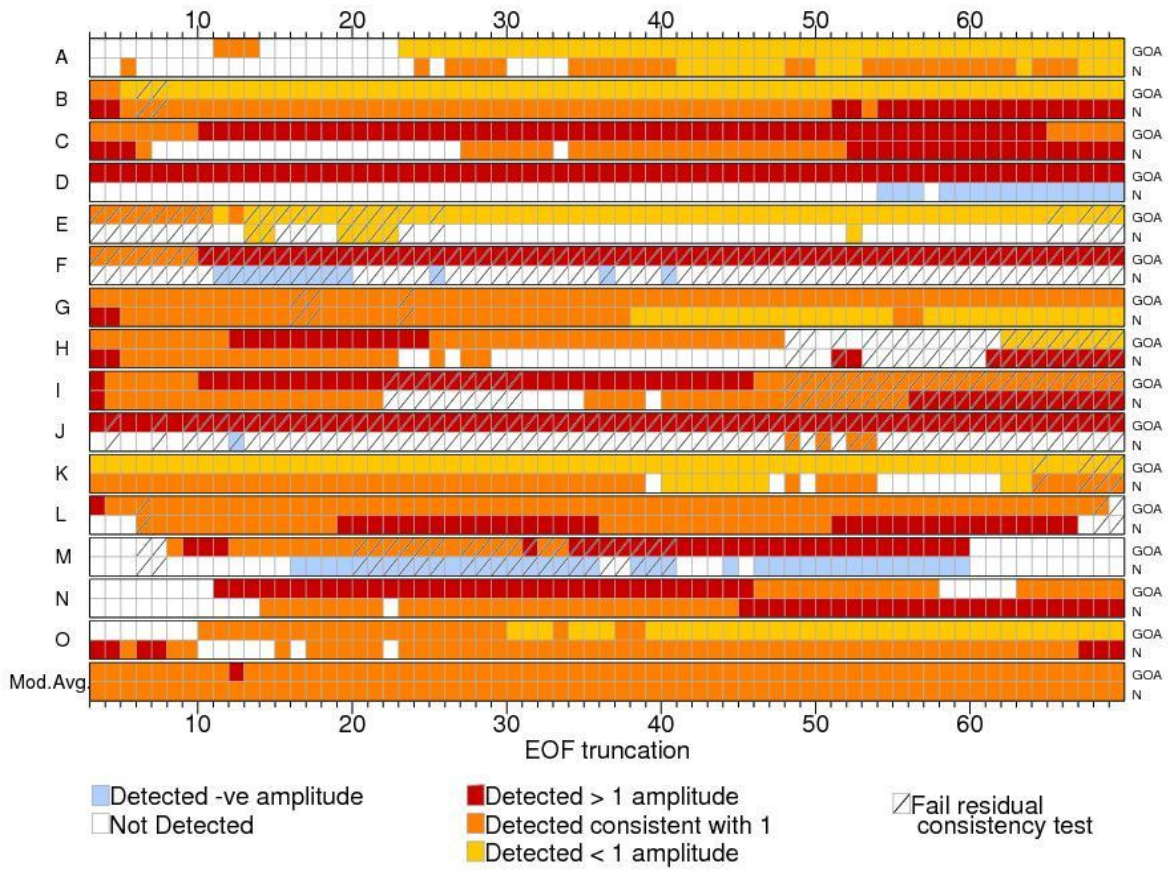

Figure S3: Summary of scaling factor sensitivity to choice of EOF truncation for optimal detection results for GOA,N analysis. See caption for Figure S1.

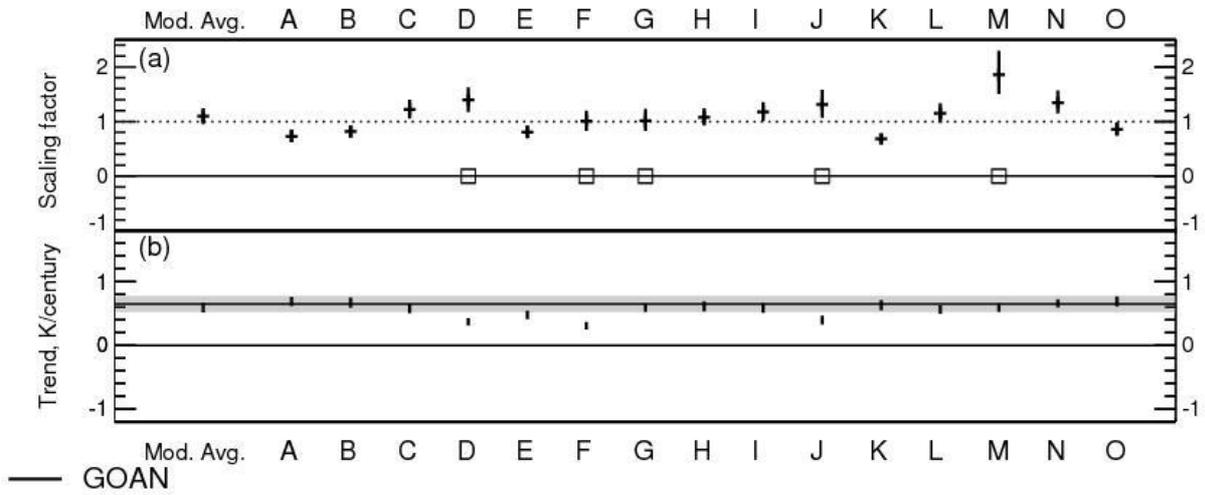

Figure S4: Optimal detection results of GOAN analysis. Scaling factors (a) and scaled trends (b) for the standard analysis of the 1906-2005 period, 10 year means and spatial meaning of spherical harmonics (T4). The EOF truncation used is 40. Squares indicate regressions where the residual fails an F-Test when compared with measure of internal variability.

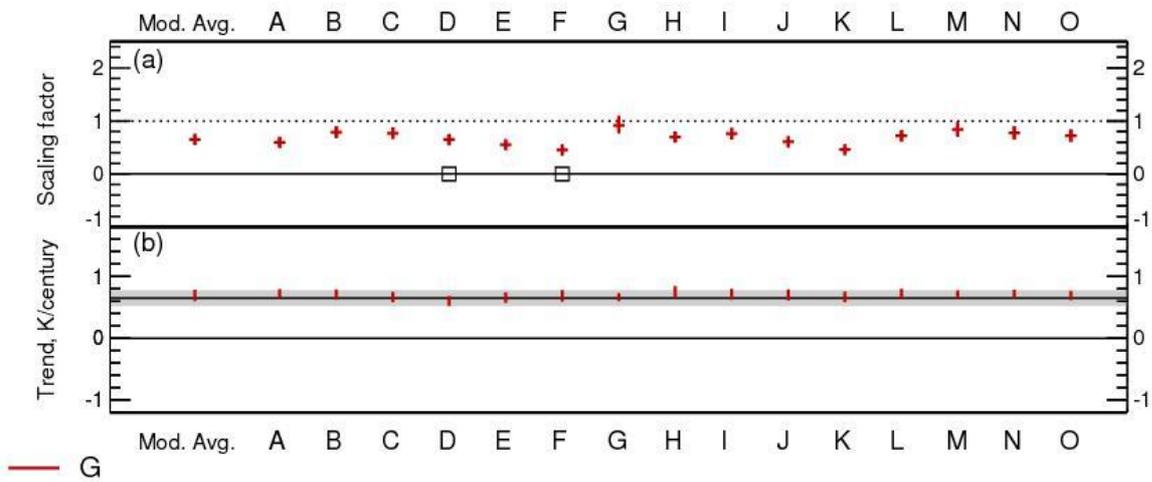

Figure S5: Optimal detection results of G analysis. See caption to Figure S4

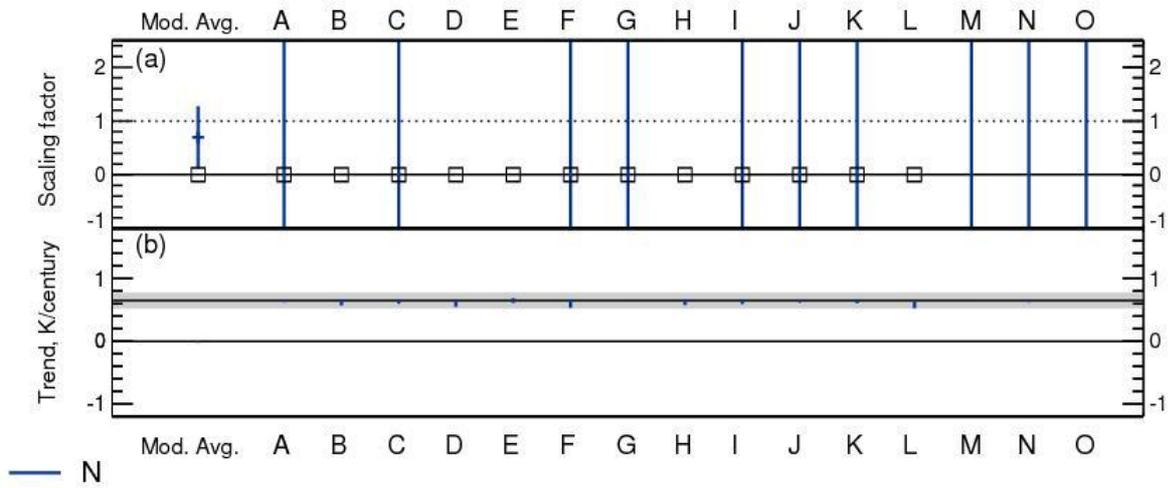

Figure S6: Optimal detection results of N analysis. See caption to Figure S4

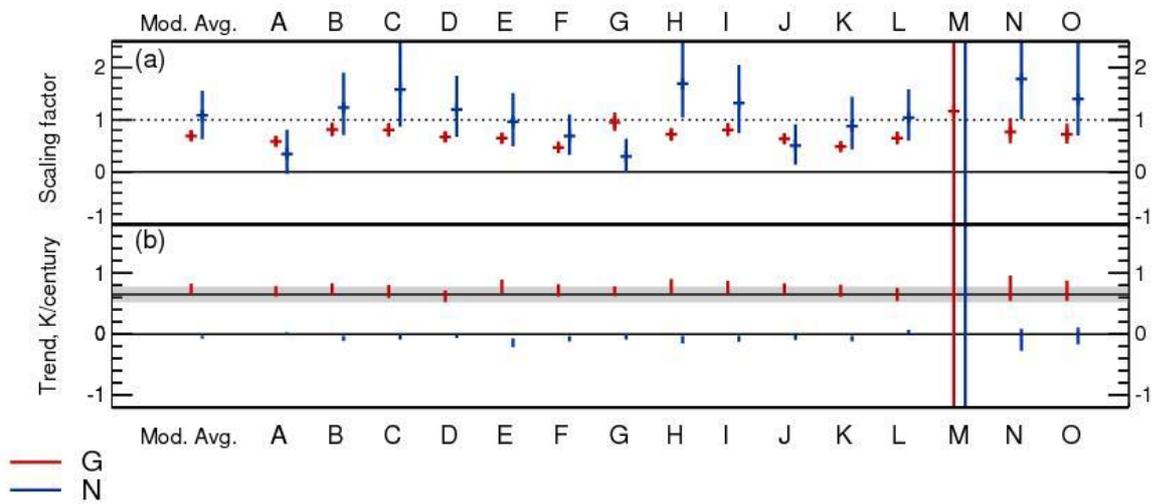

Figure S7: Optimal detection results of G,N analysis. See caption to Figure S4

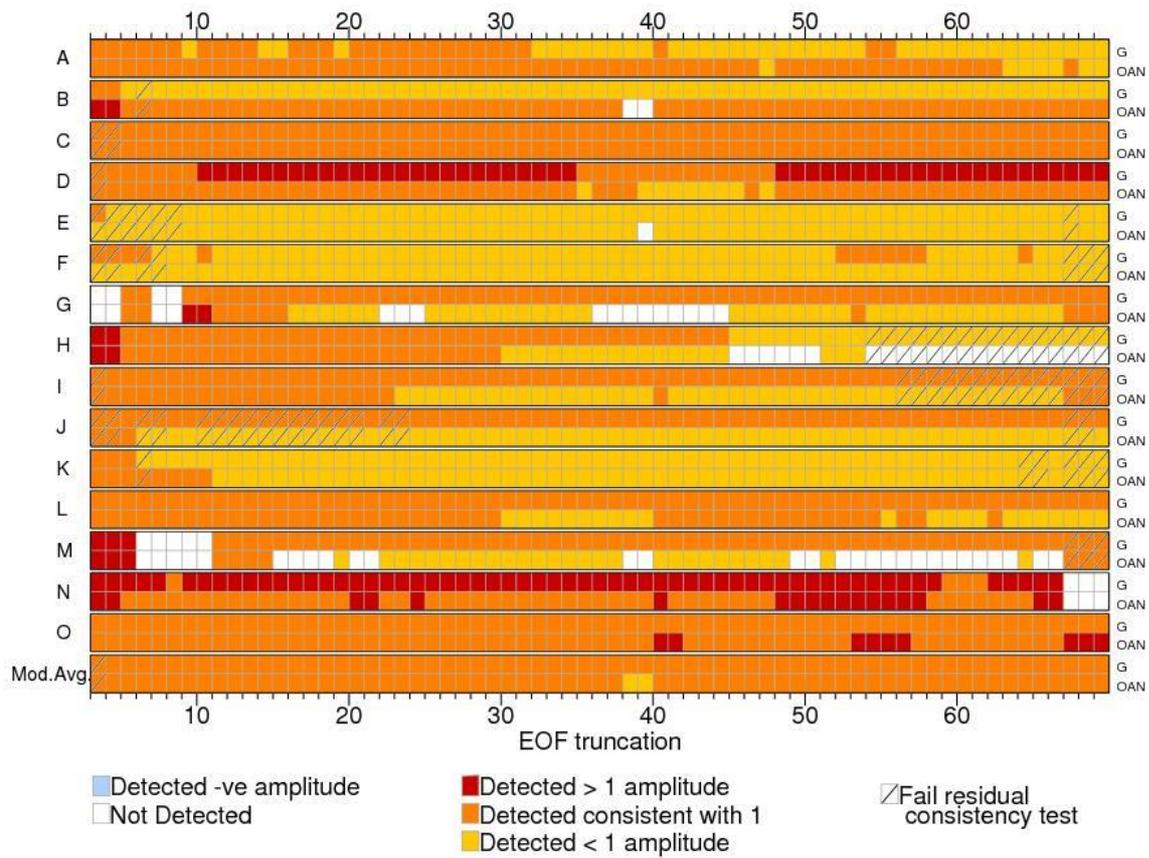

Figure S8: Summary of scaling factor sensitivity to choice of EOF truncation for optimal detection results for G,OAN analysis. See Figure S1.

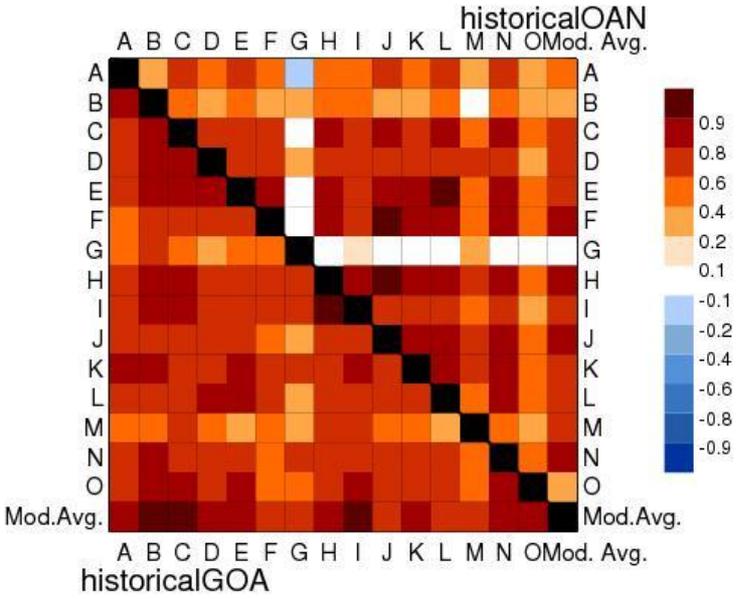

Figure S9: Area weighted spatio-temporal cross correlations between models for the 1906-2005 period and indepen- dent 10 year means. As Figure 7 in main text except for derived experiments of historicalGOA and historicalOAN.